\documentclass[12pt]{article}
\usepackage{latexsym}
\usepackage{epsfig,graphics}

\newcommand{\be}{\begin{equation}}
\newcommand{\ee}{\end{equation}}

\def\lsi{\raise0.3ex\hbox{$<$\kern-0.75em\raise-1.1ex\hbox{$\sim$}}}
\def\gsi{\raise0.3ex\hbox{$>$\kern-0.75em\raise-1.1ex\hbox{$\sim$}}}

\newcommand{\pa}{\partial}
\newcommand{\half}{{1\over 2}}

\begin{document}

\begin{titlepage}

\null\vspace{-1.0cm}

\begin{tabbing}
\` Oxford OUTP-00-53P \\
\end{tabbing}

\begin{centering}

\null\vspace{1.5cm}

{\large \bf  String models of glueballs and the spectrum of
SU(N) gauge theories in 2+1 dimensions}

\vspace{1.0cm}
Robert W. Johnson and Michael J. Teper

\vspace{0.7cm}
{\it{Theoretical Physics, University of Oxford, 1 Keble Road, \\
Oxford OX1 3NP, UK\\}}

\vspace{1.5cm}
{\bf Abstract.}
\end{centering}

\noindent
The spectrum of glueballs in 2+1 dimensions is calculated within 
an extended class of Isgur-Paton flux tube models and compared 
to lattice calculations of the low-lying SU($N\geq 2$) glueball 
mass spectrum. 
Our modifications of the model include a string curvature term and 
a new way of dealing with the short-distance cut-off. We find that 
the generic model is remarkably successful at reproducing the positive
charge conjugation, $C=+$, sector of the spectrum. The only large 
(and robust) discrepancy involves the $0^{-+}$ state, raising 
the interesting possibility that the lattice spin identification 
is mistaken and that this state is in fact $4^{-+}$. Additionally, the
Isgur-Paton model does not incorporate any mechanism for splitting 
$C=-$ from $C=+$ (in contrast to the case in 3+1 dimensions),
while the `observed' spectrum does show a substantial splitting.
We explore several modifications of the model in an attempt
to incorporate this physics in a natural way. At the qualitative
level we find that this constrains our choice to the picture
in which the $C=\pm$ splitting is driven by mixing with 
new states built on closed loops of adjoint flux. However
a detailed numerical comparison suggests that a model incorporating   
an additional direct mixing between loops of opposite orientation
is likely to work better; and that, in any case, a non-zero
curvature term will be required. We also
point out that a characteristic of any string model of glueballs
is that the SU($N\to\infty$) mass spectrum will consist of
multiple towers of states that are scaled up copies of each other. 
To test this will require a lattice mass spectrum that extends to
somewhat larger masses than currently available.  

\vspace{0.5cm}

\vfill

\end{titlepage}

\section{Introduction}
\label{sec_intro}

While it is now possible to calculate the spectrum of 
continuum non-Abelian gauge theories with some precision,
using standard lattice Monte Carlo techniques
\cite{MTrev98,MP,MTd3},
we know little about the structure of these glueballs. 
This is to be contrasted
with states containing quarks where, at least for the
low-lying spectrum, the quark model provides a
remarkably successful semi-quantitative model framework for
understanding the structure of mesons and baryons.
(Apart from the interesting cases of scalar mesons
and pseudoscalar flavour-singlet mesons.)

The glueballs of the SU(3) non-Abelian gauge theory
in 3+1 dimensions are particularly important because
the presence of such extra non-quarkonium
states in the spectrum of QCD (and in the experimental
spectrum) would provide a direct reflection of
the gauge fields in the theory. Understanding 
just how they do so (mixing, decays etc.) would
be made easier if we understood something about their
structure. Unfortunately, beyond providing some information
about glueball sizes, lattice Monte Carlo calculations
have as yet given us little insight into their structure. 
Such calculations involve connected correlators of 
several operators and while they are simple in principle it is, 
in practice, much harder to achieve sufficient statistical 
accuracy than in the corresponding mass calculations.

An alternative way to learn about the structure of glueballs
is through a reliable model -- just as the quark model 
provides us with useful information on the structure of
the low-lying mesons and baryons. To establish 
whether a glueball model is `reliable' one can compare 
the spectrum it predicts to the known spectrum (as calculated 
from the lattice). This is the approach we follow here.
There are two obvious models that one might try: constituent
gluon models, such as gluon potential
\cite{cgmodel} 
and bag models
\cite{bags},
or flux tube (string) models
\cite{IsgPat}. 
In this paper we shall confine ourselves to a study of the latter.

Models try to isolate the essential physics and neglect
everything else; thus they necessarily involve approximations. 
So one does not expect precise agreement with the known spectrum.
If we are only looking for semi-quantitative or even 
qualitative agreement, it is important to test
the model in as many relevant contexts  as possible.
One fact we can usefully use here is that string
models (and indeed bag models) can be equally motivated in 
any gauge theory that has linear confinement and hence 
string-like flux tubes. This suggests that it would be
useful to test this model not only in 3+1 dimensions
\cite{tepmor}
but also in 2+1 dimensions where non-Abelian 
gauge theories  appear to be linearly confining and detailed 
mass spectra are available 
\cite{MTd3}.  
This is what we shall do in this paper.
(We remark that recent improvements in the lattice calculation
of the D=3+1 SU(3) spectrum
\cite{MP}
warrant a complete update of the study in
\cite{tepmor}.) 

In the next section we briefly review the Isgur-Paton
flux tube model for glueballs
\cite{IsgPat}
and describe the qualitative features of the mass spectrum 
that it predicts for SU(N) gauge theories in D=2+1. 
We compare this spectrum with the `true' spectrum as calculated 
on the lattice, 
\cite{MTd3},  
which, for the reader's convenience, we summarise in 
Table~\ref{table_mlattice} borrowed from
\cite{MTd3}.
We point out where the main discrepancies and difficulties 
lie and we point out some compelling generalisations of the model.
In the subsequent section we address a major such difficulty:
how to incorporate an acceptable $C=\pm$ splitting into
the model. We finish with a summary of our results.
Finally we remark that a summary of some of our
preliminary results has appeared elsewhere
\cite{lat97},
and a much more detailed exposition appears in 
\cite{RJthesis}.
\section{The Isgur-Paton flux tube model of glueballs}
\label{sec_fluxtube}

Consider a quark and an antiquark sufficiently far apart. In
a linearly confining theory, they will be joined by a
flux tube which contributes an energy that is approximately 
proportional to its length. One can attempt to use such states 
as the basis for a flux tube model of quarkonia. The corresponding model for 
glueballs would be based on a loop of fundamental colour flux that closes
on itself. This colour singlet object contains no quarks. If we neglect its
thickness, we have a closed string of flux, and the glueball
mass spectrum is obtained by finding the energy eigenstates
of the quantised string.
This is the starting point for the Isgur-Paton model 
\cite{IsgPat}
and our consequent extensions.

We start with a closed string of colour flux in the form 
of a circle of radius $\rho$ with bare string 
tension $\sigma_b$, and hence a bare energy
\be 
E_b = 2 \pi \sigma_b \rho. 
\label{eqn_2.0}
\ee
Fluctuations about this circle are decomposed into discrete phonons of definite helicity.  These phonons carry angular momentum, and so there must be a term proportional to the total phonon number M, given by
\be 
E_{phonons} \equiv {M\over\rho}
= {1\over\rho} \sum_{m=2} m (n_m^+ + n_m^-)
\label{eqn_2.3}
\ee
added to the total energy of the state. Note that
the above sums begin with the $m=2$ mode. The $m=1$
mode is excluded in the model
\cite{IsgPat}
because infinitesimal $m=1$ fluctuations are the same as
infinitesimal translations of the circle.

As usual when one quantises over modes of all frequencies 
there is a divergent contribution to the vacuum energy.
Part of this divergent piece can be absorbed into the
renormalised string tension, $\sigma=\sigma_b+c/2\pi$. The
rest appears as a string Casimir energy. It is universal for
bosonic strings and, in the case where the strings end on 
static quarks, is called the L\"uscher term
\cite{Luscherstring}.
In the present case the string has periodic rather than fixed boundary 
conditions, just as in calculations involving Polyakov loops
\cite{Polystring},
but the coefficient differs due to the exclusion of the $m=1$ mode.
Putting all this together we can write 
the energy of the string plus its modes as
\be 
E_s = 2\pi \rho \sigma - {13\over{12\rho}} + 
{1\over\rho} \sum_{m=2} m (n_m^+ + n_m^-)
\label{eqn_2.5}
\ee
However, we know that what we have is not really a one-dimensional
string but rather a flux tube whose width will be 
$\sim 1/\surd\sigma$. For the low-lying states of interest to us 
we expect $\rho \sim  1/\surd\sigma$ and so one might expect
the simple harmonic fluctuations of the flux tube to be 
somehow suppressed. This is incorporated within the Isgur-Paton model
\cite{IsgPat}
by multiplying the contribution of the phonons to the energy
(including that of the vacuum) by a heuristic suppression 
factor that rapidly approaches unity as $\rho$ increases.
This leads to a final string energy 
\be 
E_s^M(\rho) = 2\pi \rho \sigma + 
{{M+\gamma}\over{\rho}}F(\rho)
\label{eqn_2.6}
\ee
where $\gamma=-13/12$, $M$ is defined in eqn(\ref{eqn_2.3})
and $F(\rho)$ is the factor that suppresses the string
excitations at small $\rho$. In the original model
this was chosen to be $F(\rho) = 1-e^{-f\rho}$,
where $f$ is a parameter which we would expect to be
$O(\surd\sigma)$. This form is reasonable but somewhat arbitrary; one
might ask, for example, why the string energy, $2\pi\rho\sigma$,
is not modified at small $\rho$ as well.

To quantize the string, we must take into account its motion in the radial direction as well.  In the Isgur-Paton model one identifies the conjugate momentum for the string
and writes a Schrodinger equation in the radial coordinate 
$\rho$:
\be 
\biggl\{{-9\over{16\pi\sigma}}{d^2\over{d\xi^2}}
+ E_s^M(\xi^{2/3})\biggr\}\psi(\xi) = E\psi(\xi) 
\label{eqn_2.7}
\ee
where $\xi=\rho^{3/2}$ turns out to be the natural variable to use 
here. This formalism assumes that the phonon modes are `fast'
compared to the collective radial modes and that they
can therefore be treated as providing an effective potential
for these latter modes. Clearly such an `adiabatic' 
assumption is at best approximate:
the model has only one scale, $\surd\sigma$, and so
there is no reason for the phonon fluctuations to be
$much$ faster than the collective radial fluctuations for
the low-lying part of the spectrum that will interest us.
(Indeed if one calculates
\cite{tepmor}
the low-lying spectrum one finds that
the energy splitting associated with an increment in the 
phonon number is of the same order as the splitting associated 
with an increment in the radial quantum number. This
suggests that the division into fast and slow modes is
a crude approximation at best.)

If we were in  3+1 rather than in 2+1 dimensions,
the above description of the model would change as follows.
First, rotations of the flux-loop around a diameter provide
an additional source of angular momentum, and 
equation(\ref{eqn_2.7}) acquires a corresponding angular
momentum term. In addition there are extra phonons
arising from fluctuations of the loop normal to its plane.
This doubling of modes leads to a doubling of the
value of the string Casimir energy in eqn(\ref{eqn_2.5}).

The simplest version of the Isgur-Paton model sets $F(\rho)=1$ in 
equation(\ref{eqn_2.6}) so that there is no fudge factor.  Since $\sigma$
merely sets the overall scale of the mass spectrum, the mass ratios
$m/\sqrt{\sigma}$ are then predicted with no free parameters at all.
These prediction (borrowed from reference~\cite{tepmor}) are listed in
Table~\ref{table_tepmor}.  The comparison is with $SU(N\rightarrow\infty)$, 
since in that limit, just as in the model, heavy glueballs do not decay.
The overall qualitative agreement is remarkable, with only the $0^{-+}$ 
far from its prediction in the $C=+$ sector, and motivates the more
detailed investigation of this paper.

\section{Generalising the Isgur-Paton flux tube model}
\label{sec_IPmodified}

In this section we point to several ways in which the
original Isgur-Paton model can be generalised. 
We start with the observation that one can build on
other strings than the fundamental. We then point 
out that a curvature term in the effective string
action makes an important difference. The argument
for both these extensions is compelling. We then turn
to the short-distance fudge factor $F(\rho)$ in
eqn(\ref{eqn_2.6}), point out its shortcomings and
suggest some alternatives. We leave to the next
section the important question of how to split the $C=\pm$ 
spectra in a way that is both natural in terms of the 
string model and reproduces the main features of the
observed splitting.

\subsection{Extra strings, extra states}
\label{subsec_strings}

The flux tube in the Isgur-Paton model contains flux 
in the fundamental representation; it joins charges 
that are in that representation. For SU($N\geq 4$) 
there exist charges in higher representations which
cannot be screened by virtual adjoint charges
(i.e. gluons) down to the fundamental. One can label
charges in these representations by the way they
transform under a centre gauge transformation, $z\in Z_N$.
If they acquire a factor $z^k$ we will refer to them 
as having ${N}$-ality $k$. Since gluons transform 
trivially under the centre they cannot screen the
${N}$-ality of a charge. For each such charge we
have a flux tube of a corresponding ${N}$-ality $k$,
which will possess a string tension $\sigma_k$.
We can consider a closed tube of such flux, and we can
then build a whole spectrum of glueball states on
this flux string just as we did for the fundamental,
$k=1$, string in the Isgur-Paton model. Thus as $N$
grows the spectrum will acquire extra towers of states
that are identical to the spectrum obtained with
the fundamental flux loop except that their overall
energy scale is $\sqrt{\sigma_k/\sigma_1}$ (ignoring
any mixing).

If observed, such a spectrum would be a remarkable manifestation
of the underlying string structure of glueballs. Of course
it is not guaranteed that such a spectrum actually
exists in the string picture: this will depend on the
dynamics. Consider for example the case of SU(4). If it 
happens to be the case that $\sigma_{k=2}{\not<}2\sigma_{k=1}$
then the $k=2$ flux tube can break up into two $k=1$
(fundamental) flux loops, so that the $k=2$ states are
just multi-glueball scattering states formed
out of the $k=1$ glueball states. In the SU(4) gauge
theory it is known 
\cite{blmtsu4}
that $\sigma_{k=2} \simeq 1.4\sigma_{k=1}$ in both D=2+1 and D=3+1,
and so the corresponding extra states should exist there. However the lattice
calculations have not identified enough excitations (in each
$J^{PC}$ sector) to test for the possible presence of such 
extra states, so we will ignore this potential
state replication in the remainder of this paper, apart
from pointing to its great interest for future calculations
\cite{RJthesis}.
\subsection{Curvature/elasticity}
\label{subsec_gamma}

The flux tube must have a finite thickness if it is to have a
finite energy density; presumably it will be $O(1/\surd\sigma)$.
Such a finite flux tube will presumably possess an effective elasticity.
In string language this is a curvature term. For a meson, the
curvature of the straight string joining the quarks is zero and
so a curvature term would have no effect to the order in $1/\rho$
that we are including. For a closed string, on the other hand,
the curvature is constant and integrates to a $\sim 1/\rho$
contribution. The constant of proportionality,
our effective elasticity, we denote by $\gamma_E$ and we will regard
it as an unknown free parameter.  Note that we may regard
the Casimir energy of the closed loop as simply renormalizing 
the elasticity 
\be
\gamma = \gamma_E - \frac{13}{12}
\label{eqn_2.8}
\ee
just as the $c\rho$ piece was absorbed into
a renormalisation of $\sigma$. Although there has been some discussion
\cite{KC} as to the sign such an elasticity should take, we shall 
leave $\gamma$ as a free parameter, whose value is to be determined 
by fitting the spectrum.

\subsection{Modification at short distances}
\label{subsec_f}

Since the flux tube has a finite width, a glueball will
presumably cease to look like an excited closed string
when $\rho$ is much less than that width. This is embodied 
in the Isgur-Paton model by a fudge-factor $F(\rho) = 1-e^{-f\rho}$ 
which suppresses the contribution of the string 
phonons as $\rho \to 0$, as in eqn(\ref{eqn_2.6}).
The detailed form of $F(\rho)$ is largely arbitrary,
as is the choice to suppress the phonon excitations but
not the $2\pi\rho$ string contribution. Since the spectrum
of the string model is non-singular when we set $F(\rho)=1$,
the effects of the suppression factor are not large and
the details do not matter greatly. We have calculated
the spectrum for various possibilities and we find that
as far as the $C=+$ spectrum is concerned what one
needs is a modest short-distance suppression so as to
get the $ 0^{++},2^{++}$ splitting about right, and then one
can tune the $\gamma/\rho$ contribution so as to raise
the overall spectrum to about the right level.

A quite different possibility is 
to make the string tension a function of $\rho$ rather
than to impose a fudge-factor $F(\rho)$. This is motivated
by a recent study
\cite{KST99}
of closed flux tubes in the dual Ginzburg-Landau theory.
They find that the effective string tension, $\sigma_{eff}(\rho)$,
varies with $\rho$ so as to vanish as $\rho\to 0$. One can in fact
parameterise the observed variation quite accurately using
\be 
\sigma_\mathit{eff}(\rho) = \sigma (1 - e^{-1.72 \rho}).
\label{eqn_2.46}
\ee 
One can then quantise the string model with this 
$\rho$-dependent string tension and solve for its spectrum. 
Since $\sigma_\mathit{eff}(\rho)$ appears in the mass
and hence in the kinetic energy of the loop, the quantisation
is not entirely straightforward and we leave its description
to the Appendix. (More details may be found in
\cite{RJthesis}.)
The qualitative effect of using a variable $\sigma_\mathit{eff}(\rho)$
in eqn(\ref{eqn_2.7}) is that at small $\rho$ the kinetic
energy is enhanced relative to the potential energy. This
is much the same as the effect of our fudge-factor $F(\rho)$.
However it has the advantage that there is no free parameter
(or functional form) to choose and there is no ambiguity as
to how one should apply it.

\section{Splitting ${\bf C=+}$ from ${\bf C=-}$}
\label{sec_Csplit}

For $N>2$ the flux tube carries an arrow.  In the simple Isgur-Paton model
there is no mixing between loops of opposite flux, and the resulting states
are degenerate.  Since the direction of the flux reverses under charge 
conjugation $C$, this predicts degenerate $C=\pm$ spectra, as in 
Table~\ref{table_tepmor}.
We now turn to the problem of how one might split this $C=\pm$ 
degeneracy in a way that is both natural in terms of the string 
model and reproduces the main features of the observed splitting.
We shall begin by summarising what these features are and we 
shall then consider two possible dynamical mechanisms, direct mixing
and adjoint mixing.  Additional mechanisms, such as indirect mixing and 
k-string mixing, are explored in
\cite{RJthesis}.
In each case we shall ask how well the main observed features
are reproduced.

\subsection{The observed ${\bf C=+/-}$ splitting}
\label{subsec_Cfeatures}

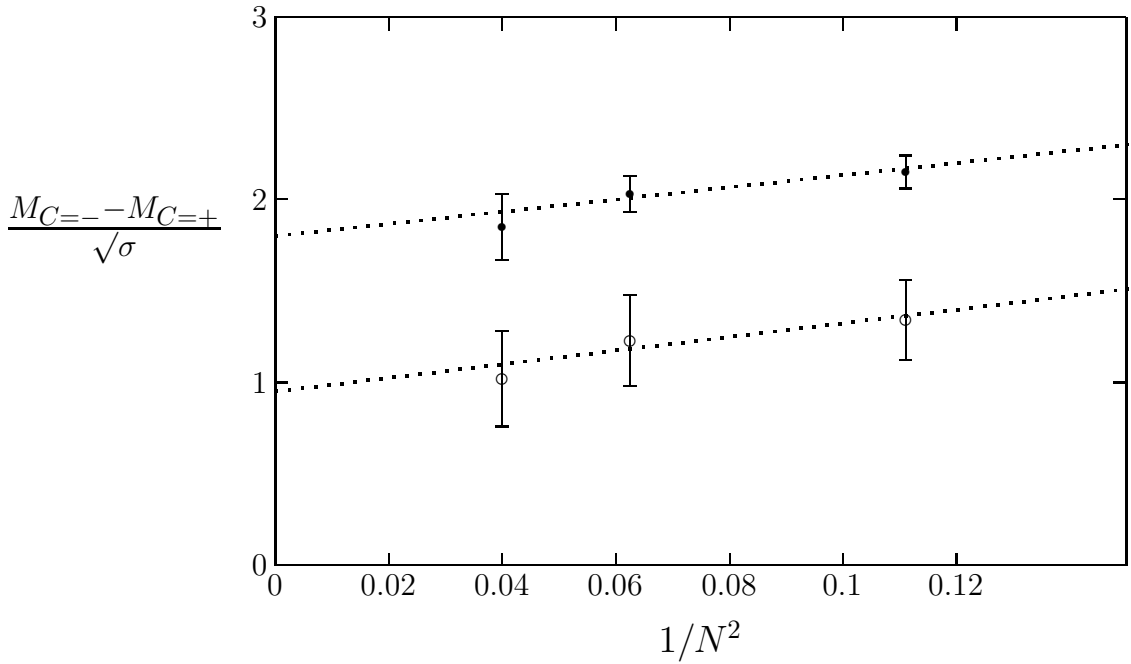
\begin	{figure}[p]
\begin	{center}
\leavevmode
\setlength{\unitlength}{0.240900pt}
\ifx\plotpoint\undefined\newsavebox{\plotpoint}\fi
\sbox{\plotpoint}{\rule[-0.200pt]{0.400pt}{0.400pt}}%
\begin{picture}(1500,900)(0,0)
\font\gnuplot=cmr10 at 12pt
\gnuplot
\sbox{\plotpoint}{\rule[-0.200pt]{0.400pt}{0.400pt}}%
\put(120.0,31.0){\rule[-0.200pt]{4.818pt}{0.400pt}}
\put(108,31){\makebox(0,0)[r]{{$0$}}}
\put(1436.0,31.0){\rule[-0.200pt]{4.818pt}{0.400pt}}
\put(120.0,318.0){\rule[-0.200pt]{4.818pt}{0.400pt}}
\put(108,318){\makebox(0,0)[r]{{$1$}}}
\put(1436.0,318.0){\rule[-0.200pt]{4.818pt}{0.400pt}}
\put(120.0,606.0){\rule[-0.200pt]{4.818pt}{0.400pt}}
\put(108,606){\makebox(0,0)[r]{{$2$}}}
\put(1436.0,606.0){\rule[-0.200pt]{4.818pt}{0.400pt}}
\put(120.0,893.0){\rule[-0.200pt]{4.818pt}{0.400pt}}
\put(108,893){\makebox(0,0)[r]{{$3$}}}
\put(1436.0,893.0){\rule[-0.200pt]{4.818pt}{0.400pt}}
\put(120.0,31.0){\rule[-0.200pt]{0.400pt}{4.818pt}}
\put(120,19){\makebox(0,0){\shortstack{\\ \\ \\ {$0$}}}}
\put(120.0,873.0){\rule[-0.200pt]{0.400pt}{4.818pt}}
\put(298.0,31.0){\rule[-0.200pt]{0.400pt}{4.818pt}}
\put(298,19){\makebox(0,0){\shortstack{\\ \\ \\ {$0.02$}}}}
\put(298.0,873.0){\rule[-0.200pt]{0.400pt}{4.818pt}}
\put(476.0,31.0){\rule[-0.200pt]{0.400pt}{4.818pt}}
\put(476,19){\makebox(0,0){\shortstack{\\ \\ \\ {$0.04$}}}}
\put(476.0,873.0){\rule[-0.200pt]{0.400pt}{4.818pt}}
\put(654.0,31.0){\rule[-0.200pt]{0.400pt}{4.818pt}}
\put(654,19){\makebox(0,0){\shortstack{\\ \\ \\ {$0.06$}}}}
\put(654.0,873.0){\rule[-0.200pt]{0.400pt}{4.818pt}}
\put(833.0,31.0){\rule[-0.200pt]{0.400pt}{4.818pt}}
\put(833,19){\makebox(0,0){\shortstack{\\ \\ \\ {$0.08$}}}}
\put(833.0,873.0){\rule[-0.200pt]{0.400pt}{4.818pt}}
\put(1011.0,31.0){\rule[-0.200pt]{0.400pt}{4.818pt}}
\put(1011,19){\makebox(0,0){\shortstack{\\ \\ \\ {$0.1$}}}}
\put(1011.0,873.0){\rule[-0.200pt]{0.400pt}{4.818pt}}
\put(1189.0,31.0){\rule[-0.200pt]{0.400pt}{4.818pt}}
\put(1189,19){\makebox(0,0){\shortstack{\\ \\ \\ {$0.12$}}}}
\put(1189.0,873.0){\rule[-0.200pt]{0.400pt}{4.818pt}}
\put(120.0,31.0){\rule[-0.200pt]{321.842pt}{0.400pt}}
\put(1456.0,31.0){\rule[-0.200pt]{0.400pt}{207.656pt}}
\put(120.0,893.0){\rule[-0.200pt]{321.842pt}{0.400pt}}
\put(-132,558){\makebox(0,0){{\Large{${{M_{C=-}-M_{C=+}}\over{\surd\sigma}}$}}}}
\put(788,-89){\makebox(0,0){{\large{$1/N^2$}}}}
\put(120.0,31.0){\rule[-0.200pt]{0.400pt}{207.656pt}}
\put(1110,649){\circle*{12}}
\put(677,614){\circle*{12}}
\put(476,563){\circle*{12}}
\put(1110.0,623.0){\rule[-0.200pt]{0.400pt}{12.527pt}}
\put(1100.0,623.0){\rule[-0.200pt]{4.818pt}{0.400pt}}
\put(1100.0,675.0){\rule[-0.200pt]{4.818pt}{0.400pt}}
\put(677.0,586.0){\rule[-0.200pt]{0.400pt}{13.731pt}}
\put(667.0,586.0){\rule[-0.200pt]{4.818pt}{0.400pt}}
\put(667.0,643.0){\rule[-0.200pt]{4.818pt}{0.400pt}}
\put(476.0,511.0){\rule[-0.200pt]{0.400pt}{24.813pt}}
\put(466.0,511.0){\rule[-0.200pt]{4.818pt}{0.400pt}}
\put(466.0,614.0){\rule[-0.200pt]{4.818pt}{0.400pt}}
\put(1110,416){\circle{18}}
\put(677,384){\circle{18}}
\put(476,324){\circle{18}}
\put(1110.0,353.0){\rule[-0.200pt]{0.400pt}{30.353pt}}
\put(1100.0,353.0){\rule[-0.200pt]{4.818pt}{0.400pt}}
\put(1100.0,479.0){\rule[-0.200pt]{4.818pt}{0.400pt}}
\put(677.0,313.0){\rule[-0.200pt]{0.400pt}{34.449pt}}
\put(667.0,313.0){\rule[-0.200pt]{4.818pt}{0.400pt}}
\put(667.0,456.0){\rule[-0.200pt]{4.818pt}{0.400pt}}
\put(476.0,249.0){\rule[-0.200pt]{0.400pt}{36.135pt}}
\put(466.0,249.0){\rule[-0.200pt]{4.818pt}{0.400pt}}
\put(466.0,399.0){\rule[-0.200pt]{4.818pt}{0.400pt}}
\sbox{\plotpoint}{\rule[-0.500pt]{1.000pt}{1.000pt}}%
\put(120,548){\usebox{\plotpoint}}
\put(120.00,548.00){\usebox{\plotpoint}}
\put(140.58,550.54){\usebox{\plotpoint}}
\multiput(147,551)(20.514,3.156){0}{\usebox{\plotpoint}}
\put(161.17,553.08){\usebox{\plotpoint}}
\put(181.87,554.61){\usebox{\plotpoint}}
\multiput(187,555)(20.547,2.935){0}{\usebox{\plotpoint}}
\put(202.46,557.11){\usebox{\plotpoint}}
\put(223.09,559.30){\usebox{\plotpoint}}
\multiput(228,560)(20.694,1.592){0}{\usebox{\plotpoint}}
\put(243.73,561.39){\usebox{\plotpoint}}
\put(264.34,563.72){\usebox{\plotpoint}}
\multiput(268,564)(20.547,2.935){0}{\usebox{\plotpoint}}
\put(284.94,566.23){\usebox{\plotpoint}}
\put(305.55,568.51){\usebox{\plotpoint}}
\multiput(309,569)(20.694,1.592){0}{\usebox{\plotpoint}}
\put(326.23,570.30){\usebox{\plotpoint}}
\put(346.83,572.67){\usebox{\plotpoint}}
\multiput(349,573)(20.703,1.479){0}{\usebox{\plotpoint}}
\put(367.47,574.69){\usebox{\plotpoint}}
\put(388.10,576.86){\usebox{\plotpoint}}
\multiput(390,577)(20.514,3.156){0}{\usebox{\plotpoint}}
\put(408.68,579.41){\usebox{\plotpoint}}
\put(429.27,581.89){\usebox{\plotpoint}}
\multiput(430,582)(20.703,1.479){0}{\usebox{\plotpoint}}
\put(449.96,583.46){\usebox{\plotpoint}}
\put(470.56,585.94){\usebox{\plotpoint}}
\multiput(471,586)(20.694,1.592){0}{\usebox{\plotpoint}}
\put(491.20,588.03){\usebox{\plotpoint}}
\multiput(498,589)(20.694,1.592){0}{\usebox{\plotpoint}}
\put(511.84,590.12){\usebox{\plotpoint}}
\put(532.44,592.57){\usebox{\plotpoint}}
\multiput(538,593)(20.547,2.935){0}{\usebox{\plotpoint}}
\put(553.03,595.08){\usebox{\plotpoint}}
\put(573.66,597.24){\usebox{\plotpoint}}
\multiput(579,598)(20.694,1.592){0}{\usebox{\plotpoint}}
\put(594.32,599.17){\usebox{\plotpoint}}
\put(614.94,601.38){\usebox{\plotpoint}}
\multiput(619,602)(20.703,1.479){0}{\usebox{\plotpoint}}
\put(635.58,603.40){\usebox{\plotpoint}}
\put(656.19,605.73){\usebox{\plotpoint}}
\multiput(660,606)(20.514,3.156){0}{\usebox{\plotpoint}}
\put(676.77,608.27){\usebox{\plotpoint}}
\put(697.38,610.60){\usebox{\plotpoint}}
\multiput(700,611)(20.703,1.479){0}{\usebox{\plotpoint}}
\put(718.06,612.31){\usebox{\plotpoint}}
\put(738.67,614.67){\usebox{\plotpoint}}
\multiput(741,615)(20.694,1.592){0}{\usebox{\plotpoint}}
\put(759.31,616.76){\usebox{\plotpoint}}
\put(779.94,618.92){\usebox{\plotpoint}}
\multiput(781,619)(20.547,2.935){0}{\usebox{\plotpoint}}
\put(800.53,621.43){\usebox{\plotpoint}}
\put(821.13,623.88){\usebox{\plotpoint}}
\multiput(822,624)(20.694,1.592){0}{\usebox{\plotpoint}}
\put(841.77,625.97){\usebox{\plotpoint}}
\multiput(849,627)(20.694,1.592){0}{\usebox{\plotpoint}}
\put(862.42,628.03){\usebox{\plotpoint}}
\put(883.05,630.09){\usebox{\plotpoint}}
\multiput(889,631)(20.703,1.479){0}{\usebox{\plotpoint}}
\put(903.70,632.11){\usebox{\plotpoint}}
\put(924.29,634.59){\usebox{\plotpoint}}
\multiput(930,635)(20.514,3.156){0}{\usebox{\plotpoint}}
\put(944.87,637.13){\usebox{\plotpoint}}
\put(965.49,639.31){\usebox{\plotpoint}}
\multiput(970,640)(20.703,1.479){0}{\usebox{\plotpoint}}
\put(986.13,641.33){\usebox{\plotpoint}}
\put(1006.74,643.70){\usebox{\plotpoint}}
\multiput(1011,644)(20.694,1.592){0}{\usebox{\plotpoint}}
\put(1027.41,645.49){\usebox{\plotpoint}}
\put(1048.03,647.77){\usebox{\plotpoint}}
\multiput(1051,648)(20.547,2.935){0}{\usebox{\plotpoint}}
\put(1068.62,650.28){\usebox{\plotpoint}}
\put(1089.24,652.61){\usebox{\plotpoint}}
\multiput(1092,653)(20.694,1.592){0}{\usebox{\plotpoint}}
\put(1109.87,654.70){\usebox{\plotpoint}}
\put(1130.50,656.88){\usebox{\plotpoint}}
\multiput(1132,657)(20.703,1.479){0}{\usebox{\plotpoint}}
\put(1151.16,658.79){\usebox{\plotpoint}}
\put(1171.79,660.91){\usebox{\plotpoint}}
\multiput(1173,661)(20.514,3.156){0}{\usebox{\plotpoint}}
\put(1192.37,663.46){\usebox{\plotpoint}}
\put(1212.96,665.99){\usebox{\plotpoint}}
\multiput(1213,666)(20.703,1.479){0}{\usebox{\plotpoint}}
\put(1233.60,668.02){\usebox{\plotpoint}}
\multiput(1240,669)(20.703,1.479){0}{\usebox{\plotpoint}}
\put(1254.24,670.04){\usebox{\plotpoint}}
\put(1274.82,672.56){\usebox{\plotpoint}}
\multiput(1281,673)(20.694,1.592){0}{\usebox{\plotpoint}}
\put(1295.51,674.22){\usebox{\plotpoint}}
\put(1316.12,676.62){\usebox{\plotpoint}}
\multiput(1321,677)(20.547,2.935){0}{\usebox{\plotpoint}}
\put(1336.71,679.13){\usebox{\plotpoint}}
\put(1357.34,681.33){\usebox{\plotpoint}}
\multiput(1362,682)(20.694,1.592){0}{\usebox{\plotpoint}}
\put(1377.98,683.43){\usebox{\plotpoint}}
\put(1398.59,685.74){\usebox{\plotpoint}}
\multiput(1402,686)(20.703,1.479){0}{\usebox{\plotpoint}}
\put(1419.26,687.50){\usebox{\plotpoint}}
\put(1439.88,689.78){\usebox{\plotpoint}}
\multiput(1443,690)(20.514,3.156){0}{\usebox{\plotpoint}}
\put(1456,692){\usebox{\plotpoint}}
\put(120,304){\usebox{\plotpoint}}
\put(120.00,304.00){\usebox{\plotpoint}}
\put(140.58,306.54){\usebox{\plotpoint}}
\multiput(147,307)(20.514,3.156){0}{\usebox{\plotpoint}}
\put(161.17,309.08){\usebox{\plotpoint}}
\put(181.80,311.20){\usebox{\plotpoint}}
\multiput(187,312)(20.547,2.935){0}{\usebox{\plotpoint}}
\put(202.35,314.10){\usebox{\plotpoint}}
\put(222.98,316.28){\usebox{\plotpoint}}
\multiput(228,317)(20.514,3.156){0}{\usebox{\plotpoint}}
\put(243.52,319.18){\usebox{\plotpoint}}
\put(264.14,321.41){\usebox{\plotpoint}}
\multiput(268,322)(20.703,1.479){0}{\usebox{\plotpoint}}
\put(284.78,323.43){\usebox{\plotpoint}}
\put(305.31,326.47){\usebox{\plotpoint}}
\multiput(309,327)(20.694,1.592){0}{\usebox{\plotpoint}}
\put(325.95,328.56){\usebox{\plotpoint}}
\put(346.48,331.61){\usebox{\plotpoint}}
\multiput(349,332)(20.703,1.479){0}{\usebox{\plotpoint}}
\put(367.12,333.63){\usebox{\plotpoint}}
\put(387.74,335.84){\usebox{\plotpoint}}
\multiput(390,336)(20.514,3.156){0}{\usebox{\plotpoint}}
\put(408.29,338.76){\usebox{\plotpoint}}
\put(428.92,340.92){\usebox{\plotpoint}}
\multiput(430,341)(20.547,2.935){0}{\usebox{\plotpoint}}
\put(449.47,343.84){\usebox{\plotpoint}}
\put(470.10,345.94){\usebox{\plotpoint}}
\multiput(471,346)(20.514,3.156){0}{\usebox{\plotpoint}}
\put(490.68,348.48){\usebox{\plotpoint}}
\multiput(498,349)(20.514,3.156){0}{\usebox{\plotpoint}}
\put(511.26,351.04){\usebox{\plotpoint}}
\put(531.86,353.53){\usebox{\plotpoint}}
\multiput(538,354)(20.547,2.935){0}{\usebox{\plotpoint}}
\put(552.45,356.07){\usebox{\plotpoint}}
\put(573.04,358.57){\usebox{\plotpoint}}
\multiput(579,359)(20.514,3.156){0}{\usebox{\plotpoint}}
\put(593.62,361.12){\usebox{\plotpoint}}
\put(614.25,363.27){\usebox{\plotpoint}}
\multiput(619,364)(20.547,2.935){0}{\usebox{\plotpoint}}
\put(634.80,366.14){\usebox{\plotpoint}}
\put(655.43,368.35){\usebox{\plotpoint}}
\multiput(660,369)(20.514,3.156){0}{\usebox{\plotpoint}}
\put(675.97,371.21){\usebox{\plotpoint}}
\put(696.59,373.48){\usebox{\plotpoint}}
\multiput(700,374)(20.703,1.479){0}{\usebox{\plotpoint}}
\put(717.23,375.50){\usebox{\plotpoint}}
\put(737.76,378.54){\usebox{\plotpoint}}
\multiput(741,379)(20.694,1.592){0}{\usebox{\plotpoint}}
\put(758.40,380.63){\usebox{\plotpoint}}
\put(778.93,383.68){\usebox{\plotpoint}}
\multiput(781,384)(20.703,1.479){0}{\usebox{\plotpoint}}
\put(799.57,385.70){\usebox{\plotpoint}}
\put(820.20,387.87){\usebox{\plotpoint}}
\multiput(822,388)(20.514,3.156){0}{\usebox{\plotpoint}}
\put(840.74,390.82){\usebox{\plotpoint}}
\put(861.37,392.95){\usebox{\plotpoint}}
\multiput(862,393)(20.547,2.935){0}{\usebox{\plotpoint}}
\put(881.91,395.91){\usebox{\plotpoint}}
\put(902.55,397.97){\usebox{\plotpoint}}
\multiput(903,398)(20.514,3.156){0}{\usebox{\plotpoint}}
\put(923.13,400.51){\usebox{\plotpoint}}
\multiput(930,401)(20.514,3.156){0}{\usebox{\plotpoint}}
\put(943.71,403.10){\usebox{\plotpoint}}
\put(964.31,405.56){\usebox{\plotpoint}}
\multiput(970,406)(20.547,2.935){0}{\usebox{\plotpoint}}
\put(984.90,408.14){\usebox{\plotpoint}}
\put(1005.49,410.61){\usebox{\plotpoint}}
\multiput(1011,411)(20.514,3.156){0}{\usebox{\plotpoint}}
\put(1026.07,413.15){\usebox{\plotpoint}}
\put(1046.70,415.34){\usebox{\plotpoint}}
\multiput(1051,416)(20.547,2.935){0}{\usebox{\plotpoint}}
\put(1067.25,418.17){\usebox{\plotpoint}}
\put(1087.87,420.41){\usebox{\plotpoint}}
\multiput(1092,421)(20.514,3.156){0}{\usebox{\plotpoint}}
\put(1108.43,423.24){\usebox{\plotpoint}}
\put(1129.04,425.54){\usebox{\plotpoint}}
\multiput(1132,426)(20.703,1.479){0}{\usebox{\plotpoint}}
\put(1149.68,427.57){\usebox{\plotpoint}}
\put(1170.21,430.60){\usebox{\plotpoint}}
\multiput(1173,431)(20.694,1.592){0}{\usebox{\plotpoint}}
\put(1190.85,432.69){\usebox{\plotpoint}}
\put(1211.38,435.75){\usebox{\plotpoint}}
\multiput(1213,436)(20.703,1.479){0}{\usebox{\plotpoint}}
\put(1232.02,437.77){\usebox{\plotpoint}}
\put(1252.65,439.90){\usebox{\plotpoint}}
\multiput(1254,440)(20.514,3.156){0}{\usebox{\plotpoint}}
\put(1273.19,442.88){\usebox{\plotpoint}}
\put(1293.82,444.99){\usebox{\plotpoint}}
\multiput(1294,445)(20.547,2.935){0}{\usebox{\plotpoint}}
\put(1314.36,447.98){\usebox{\plotpoint}}
\multiput(1321,449)(20.703,1.479){0}{\usebox{\plotpoint}}
\put(1335.00,450.00){\usebox{\plotpoint}}
\put(1355.59,452.54){\usebox{\plotpoint}}
\multiput(1362,453)(20.514,3.156){0}{\usebox{\plotpoint}}
\put(1376.16,455.17){\usebox{\plotpoint}}
\put(1396.76,457.60){\usebox{\plotpoint}}
\multiput(1402,458)(20.547,2.935){0}{\usebox{\plotpoint}}
\put(1417.35,460.21){\usebox{\plotpoint}}
\put(1437.94,462.64){\usebox{\plotpoint}}
\multiput(1443,463)(20.514,3.156){0}{\usebox{\plotpoint}}
\put(1456,465){\usebox{\plotpoint}}
\end{picture}

\end	{center}
\vskip 0.15in
\caption{Some of the observed SU($N$) $C=\pm$ splittings plotted
versus $1/N^2$: the mass difference between the $0^{--}$ 
and the $0^{++}$
($\bullet$) and that between the $2^{\pm -}$ and the $2^{\pm +}$
($\circ$). As $N \to \infty$ the dependence is expected to
be linear in $1/N^2$, i.e. like the straight lines added to the 
plot to guide the eye.}
\label{fig_Csplit}
\end 	{figure}

From the masses listed in Table~\ref{table_mlattice}
we see that the ${C=+/-}$ splitting possesses
the following qualitative features:

{\noindent}$\bullet$ In Fig.\ref{fig_Csplit} we plot two
examples of the ${C=\pm}$ splitting, as a
function of $1/N^2$. This is a natural variable to use
since the leading corrections to the large-$N$ limit
are expected to be $O(1/N^2)$
\cite{largeN}.
We infer from this plot (and from similar plots of
other states) that the splitting remains non-zero
in the $N=\infty$ limit; it is a leading order effect.

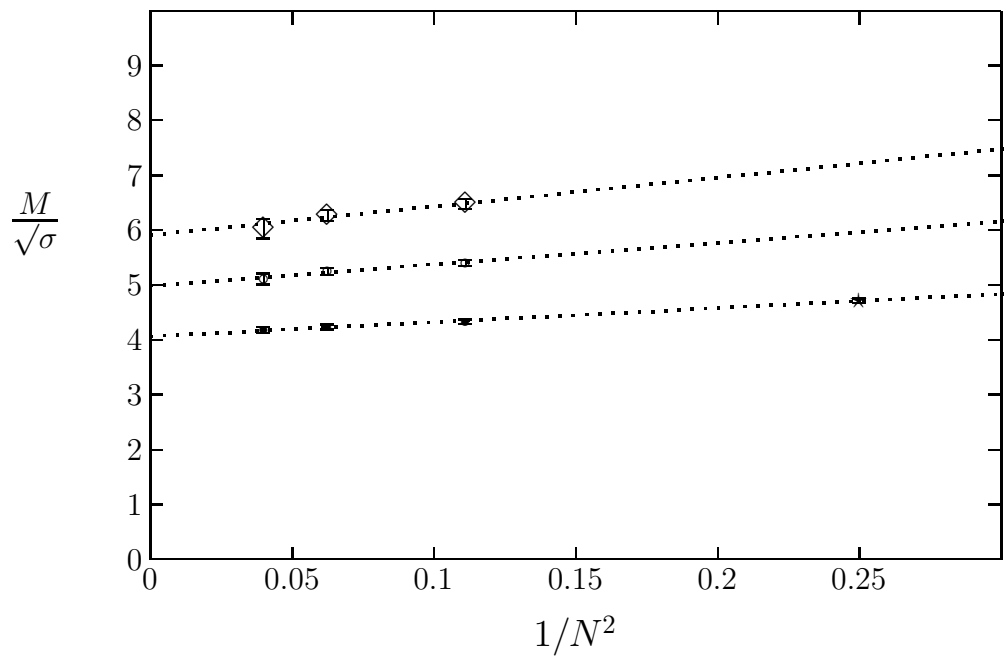
\begin	{figure}[p]
\begin	{center}
\leavevmode
\setlength{\unitlength}{0.240900pt}
\ifx\plotpoint\undefined\newsavebox{\plotpoint}\fi
\sbox{\plotpoint}{\rule[-0.200pt]{0.400pt}{0.400pt}}%
\begin{picture}(1500,900)(0,0)
\font\gnuplot=cmr10 at 12pt
\gnuplot
\sbox{\plotpoint}{\rule[-0.200pt]{0.400pt}{0.400pt}}%
\put(120.0,31.0){\rule[-0.200pt]{4.818pt}{0.400pt}}
\put(108,31){\makebox(0,0)[r]{{$0$}}}
\put(1436.0,31.0){\rule[-0.200pt]{4.818pt}{0.400pt}}
\put(120.0,117.0){\rule[-0.200pt]{4.818pt}{0.400pt}}
\put(108,117){\makebox(0,0)[r]{{$1$}}}
\put(1436.0,117.0){\rule[-0.200pt]{4.818pt}{0.400pt}}
\put(120.0,203.0){\rule[-0.200pt]{4.818pt}{0.400pt}}
\put(108,203){\makebox(0,0)[r]{{$2$}}}
\put(1436.0,203.0){\rule[-0.200pt]{4.818pt}{0.400pt}}
\put(120.0,290.0){\rule[-0.200pt]{4.818pt}{0.400pt}}
\put(108,290){\makebox(0,0)[r]{{$3$}}}
\put(1436.0,290.0){\rule[-0.200pt]{4.818pt}{0.400pt}}
\put(120.0,376.0){\rule[-0.200pt]{4.818pt}{0.400pt}}
\put(108,376){\makebox(0,0)[r]{{$4$}}}
\put(1436.0,376.0){\rule[-0.200pt]{4.818pt}{0.400pt}}
\put(120.0,462.0){\rule[-0.200pt]{4.818pt}{0.400pt}}
\put(108,462){\makebox(0,0)[r]{{$5$}}}
\put(1436.0,462.0){\rule[-0.200pt]{4.818pt}{0.400pt}}
\put(120.0,548.0){\rule[-0.200pt]{4.818pt}{0.400pt}}
\put(108,548){\makebox(0,0)[r]{{$6$}}}
\put(1436.0,548.0){\rule[-0.200pt]{4.818pt}{0.400pt}}
\put(120.0,634.0){\rule[-0.200pt]{4.818pt}{0.400pt}}
\put(108,634){\makebox(0,0)[r]{{$7$}}}
\put(1436.0,634.0){\rule[-0.200pt]{4.818pt}{0.400pt}}
\put(120.0,721.0){\rule[-0.200pt]{4.818pt}{0.400pt}}
\put(108,721){\makebox(0,0)[r]{{$8$}}}
\put(1436.0,721.0){\rule[-0.200pt]{4.818pt}{0.400pt}}
\put(120.0,807.0){\rule[-0.200pt]{4.818pt}{0.400pt}}
\put(108,807){\makebox(0,0)[r]{{$9$}}}
\put(1436.0,807.0){\rule[-0.200pt]{4.818pt}{0.400pt}}
\put(120.0,31.0){\rule[-0.200pt]{0.400pt}{4.818pt}}
\put(120,19){\makebox(0,0){\shortstack{\\ \\ \\ {$0$}}}}
\put(120.0,873.0){\rule[-0.200pt]{0.400pt}{4.818pt}}
\put(343.0,31.0){\rule[-0.200pt]{0.400pt}{4.818pt}}
\put(343,19){\makebox(0,0){\shortstack{\\ \\ \\ {$0.05$}}}}
\put(343.0,873.0){\rule[-0.200pt]{0.400pt}{4.818pt}}
\put(565.0,31.0){\rule[-0.200pt]{0.400pt}{4.818pt}}
\put(565,19){\makebox(0,0){\shortstack{\\ \\ \\ {$0.1$}}}}
\put(565.0,873.0){\rule[-0.200pt]{0.400pt}{4.818pt}}
\put(788.0,31.0){\rule[-0.200pt]{0.400pt}{4.818pt}}
\put(788,19){\makebox(0,0){\shortstack{\\ \\ \\ {$0.15$}}}}
\put(788.0,873.0){\rule[-0.200pt]{0.400pt}{4.818pt}}
\put(1011.0,31.0){\rule[-0.200pt]{0.400pt}{4.818pt}}
\put(1011,19){\makebox(0,0){\shortstack{\\ \\ \\ {$0.2$}}}}
\put(1011.0,873.0){\rule[-0.200pt]{0.400pt}{4.818pt}}
\put(1233.0,31.0){\rule[-0.200pt]{0.400pt}{4.818pt}}
\put(1233,19){\makebox(0,0){\shortstack{\\ \\ \\ {$0.25$}}}}
\put(1233.0,873.0){\rule[-0.200pt]{0.400pt}{4.818pt}}
\put(120.0,31.0){\rule[-0.200pt]{321.842pt}{0.400pt}}
\put(1456.0,31.0){\rule[-0.200pt]{0.400pt}{207.656pt}}
\put(120.0,893.0){\rule[-0.200pt]{321.842pt}{0.400pt}}
\put(-60,558){\makebox(0,0){{\Large{${{M}\over{\surd\sigma}}$}}}}
\put(788,-89){\makebox(0,0){{\large{$1/N^2$}}}}
\put(120.0,31.0){\rule[-0.200pt]{0.400pt}{207.656pt}}
\put(1233,438){\makebox(0,0){$\star$}}
\put(1233.0,434.0){\rule[-0.200pt]{0.400pt}{1.686pt}}
\put(1223.0,434.0){\rule[-0.200pt]{4.818pt}{0.400pt}}
\put(1223.0,441.0){\rule[-0.200pt]{4.818pt}{0.400pt}}
\put(615,404){\circle*{12}}
\put(398,396){\circle*{12}}
\put(298,392){\circle*{12}}
\put(615.0,401.0){\rule[-0.200pt]{0.400pt}{1.686pt}}
\put(605.0,401.0){\rule[-0.200pt]{4.818pt}{0.400pt}}
\put(605.0,408.0){\rule[-0.200pt]{4.818pt}{0.400pt}}
\put(398.0,392.0){\rule[-0.200pt]{0.400pt}{1.927pt}}
\put(388.0,392.0){\rule[-0.200pt]{4.818pt}{0.400pt}}
\put(388.0,400.0){\rule[-0.200pt]{4.818pt}{0.400pt}}
\put(298.0,387.0){\rule[-0.200pt]{0.400pt}{2.168pt}}
\put(288.0,387.0){\rule[-0.200pt]{4.818pt}{0.400pt}}
\put(288.0,396.0){\rule[-0.200pt]{4.818pt}{0.400pt}}
\put(615,496){\circle{12}}
\put(398,484){\circle{12}}
\put(298,471){\circle{12}}
\put(615.0,492.0){\rule[-0.200pt]{0.400pt}{2.168pt}}
\put(605.0,492.0){\rule[-0.200pt]{4.818pt}{0.400pt}}
\put(605.0,501.0){\rule[-0.200pt]{4.818pt}{0.400pt}}
\put(398.0,478.0){\rule[-0.200pt]{0.400pt}{2.650pt}}
\put(388.0,478.0){\rule[-0.200pt]{4.818pt}{0.400pt}}
\put(388.0,489.0){\rule[-0.200pt]{4.818pt}{0.400pt}}
\put(298.0,463.0){\rule[-0.200pt]{0.400pt}{4.095pt}}
\put(288.0,463.0){\rule[-0.200pt]{4.818pt}{0.400pt}}
\put(288.0,480.0){\rule[-0.200pt]{4.818pt}{0.400pt}}
\put(615,590){\raisebox{-.8pt}{\makebox(0,0){$\Diamond$}}}
\put(398,571){\raisebox{-.8pt}{\makebox(0,0){$\Diamond$}}}
\put(298,551){\raisebox{-.8pt}{\makebox(0,0){$\Diamond$}}}
\put(615.0,582.0){\rule[-0.200pt]{0.400pt}{3.613pt}}
\put(605.0,582.0){\rule[-0.200pt]{4.818pt}{0.400pt}}
\put(605.0,597.0){\rule[-0.200pt]{4.818pt}{0.400pt}}
\put(398.0,563.0){\rule[-0.200pt]{0.400pt}{4.095pt}}
\put(388.0,563.0){\rule[-0.200pt]{4.818pt}{0.400pt}}
\put(388.0,580.0){\rule[-0.200pt]{4.818pt}{0.400pt}}
\put(298.0,535.0){\rule[-0.200pt]{0.400pt}{7.468pt}}
\put(288.0,535.0){\rule[-0.200pt]{4.818pt}{0.400pt}}
\put(288.0,566.0){\rule[-0.200pt]{4.818pt}{0.400pt}}
\sbox{\plotpoint}{\rule[-0.500pt]{1.000pt}{1.000pt}}%
\put(120,381){\usebox{\plotpoint}}
\put(120.00,381.00){\usebox{\plotpoint}}
\put(140.70,382.55){\usebox{\plotpoint}}
\multiput(147,383)(20.756,0.000){0}{\usebox{\plotpoint}}
\put(161.43,383.10){\usebox{\plotpoint}}
\put(182.13,384.63){\usebox{\plotpoint}}
\multiput(187,385)(20.756,0.000){0}{\usebox{\plotpoint}}
\put(202.87,385.14){\usebox{\plotpoint}}
\put(223.57,386.68){\usebox{\plotpoint}}
\multiput(228,387)(20.756,0.000){0}{\usebox{\plotpoint}}
\put(244.30,387.24){\usebox{\plotpoint}}
\put(265.00,388.77){\usebox{\plotpoint}}
\multiput(268,389)(20.756,0.000){0}{\usebox{\plotpoint}}
\put(285.74,389.29){\usebox{\plotpoint}}
\put(306.44,390.82){\usebox{\plotpoint}}
\multiput(309,391)(20.694,1.592){0}{\usebox{\plotpoint}}
\put(327.15,392.00){\usebox{\plotpoint}}
\put(347.87,392.91){\usebox{\plotpoint}}
\multiput(349,393)(20.703,1.479){0}{\usebox{\plotpoint}}
\put(368.58,394.00){\usebox{\plotpoint}}
\put(389.30,394.95){\usebox{\plotpoint}}
\multiput(390,395)(20.694,1.592){0}{\usebox{\plotpoint}}
\put(410.02,396.00){\usebox{\plotpoint}}
\multiput(417,396)(20.694,1.592){0}{\usebox{\plotpoint}}
\put(430.74,397.05){\usebox{\plotpoint}}
\put(451.46,398.00){\usebox{\plotpoint}}
\multiput(457,398)(20.703,1.479){0}{\usebox{\plotpoint}}
\put(472.17,399.09){\usebox{\plotpoint}}
\put(492.89,400.00){\usebox{\plotpoint}}
\multiput(498,400)(20.694,1.592){0}{\usebox{\plotpoint}}
\put(513.60,401.19){\usebox{\plotpoint}}
\put(534.33,402.00){\usebox{\plotpoint}}
\multiput(538,402)(20.703,1.479){0}{\usebox{\plotpoint}}
\put(555.04,403.23){\usebox{\plotpoint}}
\put(575.77,404.00){\usebox{\plotpoint}}
\multiput(579,404)(20.694,1.592){0}{\usebox{\plotpoint}}
\put(596.47,405.32){\usebox{\plotpoint}}
\put(617.20,406.00){\usebox{\plotpoint}}
\multiput(619,406)(20.703,1.479){0}{\usebox{\plotpoint}}
\put(637.91,407.38){\usebox{\plotpoint}}
\put(658.64,408.00){\usebox{\plotpoint}}
\multiput(660,408)(20.694,1.592){0}{\usebox{\plotpoint}}
\put(679.34,409.45){\usebox{\plotpoint}}
\multiput(687,410)(20.756,0.000){0}{\usebox{\plotpoint}}
\put(700.08,410.01){\usebox{\plotpoint}}
\put(720.78,411.52){\usebox{\plotpoint}}
\multiput(727,412)(20.756,0.000){0}{\usebox{\plotpoint}}
\put(741.51,412.04){\usebox{\plotpoint}}
\put(762.21,413.59){\usebox{\plotpoint}}
\multiput(768,414)(20.756,0.000){0}{\usebox{\plotpoint}}
\put(782.95,414.14){\usebox{\plotpoint}}
\put(803.65,415.67){\usebox{\plotpoint}}
\multiput(808,416)(20.756,0.000){0}{\usebox{\plotpoint}}
\put(824.38,416.18){\usebox{\plotpoint}}
\put(845.08,417.72){\usebox{\plotpoint}}
\multiput(849,418)(20.756,0.000){0}{\usebox{\plotpoint}}
\put(865.82,418.27){\usebox{\plotpoint}}
\put(886.52,419.81){\usebox{\plotpoint}}
\multiput(889,420)(20.756,0.000){0}{\usebox{\plotpoint}}
\put(907.25,420.33){\usebox{\plotpoint}}
\put(927.95,421.85){\usebox{\plotpoint}}
\multiput(930,422)(20.694,1.592){0}{\usebox{\plotpoint}}
\put(948.66,423.00){\usebox{\plotpoint}}
\put(969.38,423.95){\usebox{\plotpoint}}
\multiput(970,424)(20.703,1.479){0}{\usebox{\plotpoint}}
\put(990.10,425.00){\usebox{\plotpoint}}
\put(1010.82,425.99){\usebox{\plotpoint}}
\multiput(1011,426)(20.694,1.592){0}{\usebox{\plotpoint}}
\put(1031.54,427.00){\usebox{\plotpoint}}
\multiput(1038,427)(20.694,1.592){0}{\usebox{\plotpoint}}
\put(1052.25,428.09){\usebox{\plotpoint}}
\put(1072.97,429.00){\usebox{\plotpoint}}
\multiput(1078,429)(20.703,1.479){0}{\usebox{\plotpoint}}
\put(1093.69,430.13){\usebox{\plotpoint}}
\put(1114.41,431.00){\usebox{\plotpoint}}
\multiput(1119,431)(20.694,1.592){0}{\usebox{\plotpoint}}
\put(1135.12,432.22){\usebox{\plotpoint}}
\put(1155.85,433.00){\usebox{\plotpoint}}
\multiput(1159,433)(20.703,1.479){0}{\usebox{\plotpoint}}
\put(1176.56,434.27){\usebox{\plotpoint}}
\put(1197.28,435.00){\usebox{\plotpoint}}
\multiput(1200,435)(20.694,1.592){0}{\usebox{\plotpoint}}
\put(1217.99,436.36){\usebox{\plotpoint}}
\put(1238.72,437.00){\usebox{\plotpoint}}
\multiput(1240,437)(20.703,1.479){0}{\usebox{\plotpoint}}
\put(1259.42,438.42){\usebox{\plotpoint}}
\put(1280.16,439.00){\usebox{\plotpoint}}
\multiput(1281,439)(20.694,1.592){0}{\usebox{\plotpoint}}
\put(1300.86,440.49){\usebox{\plotpoint}}
\multiput(1308,441)(20.756,0.000){0}{\usebox{\plotpoint}}
\put(1321.59,441.04){\usebox{\plotpoint}}
\put(1342.29,442.56){\usebox{\plotpoint}}
\multiput(1348,443)(20.756,0.000){0}{\usebox{\plotpoint}}
\put(1363.03,443.08){\usebox{\plotpoint}}
\put(1383.73,444.62){\usebox{\plotpoint}}
\multiput(1389,445)(20.756,0.000){0}{\usebox{\plotpoint}}
\put(1404.46,445.18){\usebox{\plotpoint}}
\put(1425.16,446.70){\usebox{\plotpoint}}
\multiput(1429,447)(20.756,0.000){0}{\usebox{\plotpoint}}
\put(1445.90,447.22){\usebox{\plotpoint}}
\put(1456,448){\usebox{\plotpoint}}
\put(120,461){\usebox{\plotpoint}}
\put(120.00,461.00){\usebox{\plotpoint}}
\put(140.70,462.55){\usebox{\plotpoint}}
\multiput(147,463)(20.694,1.592){0}{\usebox{\plotpoint}}
\put(161.39,464.10){\usebox{\plotpoint}}
\put(182.09,465.62){\usebox{\plotpoint}}
\multiput(187,466)(20.703,1.479){0}{\usebox{\plotpoint}}
\put(202.79,467.14){\usebox{\plotpoint}}
\put(223.49,468.68){\usebox{\plotpoint}}
\multiput(228,469)(20.694,1.592){0}{\usebox{\plotpoint}}
\put(244.19,470.23){\usebox{\plotpoint}}
\put(264.89,471.76){\usebox{\plotpoint}}
\multiput(268,472)(20.703,1.479){0}{\usebox{\plotpoint}}
\put(285.59,473.28){\usebox{\plotpoint}}
\put(306.29,474.81){\usebox{\plotpoint}}
\multiput(309,475)(20.694,1.592){0}{\usebox{\plotpoint}}
\put(326.99,476.36){\usebox{\plotpoint}}
\put(347.68,477.90){\usebox{\plotpoint}}
\multiput(349,478)(20.703,1.479){0}{\usebox{\plotpoint}}
\put(368.38,479.41){\usebox{\plotpoint}}
\put(389.08,480.93){\usebox{\plotpoint}}
\multiput(390,481)(20.694,1.592){0}{\usebox{\plotpoint}}
\put(409.78,482.48){\usebox{\plotpoint}}
\multiput(417,483)(20.694,1.592){0}{\usebox{\plotpoint}}
\put(430.48,484.03){\usebox{\plotpoint}}
\put(451.18,485.55){\usebox{\plotpoint}}
\multiput(457,486)(20.703,1.479){0}{\usebox{\plotpoint}}
\put(471.88,487.07){\usebox{\plotpoint}}
\put(492.51,489.22){\usebox{\plotpoint}}
\multiput(498,490)(20.694,1.592){0}{\usebox{\plotpoint}}
\put(513.17,491.15){\usebox{\plotpoint}}
\put(533.87,492.68){\usebox{\plotpoint}}
\multiput(538,493)(20.703,1.479){0}{\usebox{\plotpoint}}
\put(554.57,494.20){\usebox{\plotpoint}}
\put(575.26,495.73){\usebox{\plotpoint}}
\multiput(579,496)(20.694,1.592){0}{\usebox{\plotpoint}}
\put(595.96,497.28){\usebox{\plotpoint}}
\put(616.66,498.82){\usebox{\plotpoint}}
\multiput(619,499)(20.703,1.479){0}{\usebox{\plotpoint}}
\put(637.36,500.34){\usebox{\plotpoint}}
\put(658.06,501.86){\usebox{\plotpoint}}
\multiput(660,502)(20.694,1.592){0}{\usebox{\plotpoint}}
\put(678.76,503.41){\usebox{\plotpoint}}
\put(699.46,504.96){\usebox{\plotpoint}}
\multiput(700,505)(20.703,1.479){0}{\usebox{\plotpoint}}
\put(720.16,506.47){\usebox{\plotpoint}}
\put(740.86,507.99){\usebox{\plotpoint}}
\multiput(741,508)(20.694,1.592){0}{\usebox{\plotpoint}}
\put(761.55,509.54){\usebox{\plotpoint}}
\multiput(768,510)(20.694,1.592){0}{\usebox{\plotpoint}}
\put(782.25,511.09){\usebox{\plotpoint}}
\put(802.95,512.61){\usebox{\plotpoint}}
\multiput(808,513)(20.703,1.479){0}{\usebox{\plotpoint}}
\put(823.65,514.13){\usebox{\plotpoint}}
\put(844.35,515.67){\usebox{\plotpoint}}
\multiput(849,516)(20.694,1.592){0}{\usebox{\plotpoint}}
\put(865.05,517.22){\usebox{\plotpoint}}
\put(885.74,518.75){\usebox{\plotpoint}}
\multiput(889,519)(20.703,1.479){0}{\usebox{\plotpoint}}
\put(906.44,520.26){\usebox{\plotpoint}}
\put(927.14,521.80){\usebox{\plotpoint}}
\multiput(930,522)(20.694,1.592){0}{\usebox{\plotpoint}}
\put(947.84,523.35){\usebox{\plotpoint}}
\put(968.54,524.89){\usebox{\plotpoint}}
\multiput(970,525)(20.703,1.479){0}{\usebox{\plotpoint}}
\put(989.24,526.40){\usebox{\plotpoint}}
\put(1009.94,527.92){\usebox{\plotpoint}}
\multiput(1011,528)(20.694,1.592){0}{\usebox{\plotpoint}}
\put(1030.64,529.47){\usebox{\plotpoint}}
\multiput(1038,530)(20.694,1.592){0}{\usebox{\plotpoint}}
\put(1051.33,531.02){\usebox{\plotpoint}}
\put(1072.03,532.54){\usebox{\plotpoint}}
\multiput(1078,533)(20.703,1.479){0}{\usebox{\plotpoint}}
\put(1092.73,534.06){\usebox{\plotpoint}}
\put(1113.37,536.20){\usebox{\plotpoint}}
\multiput(1119,537)(20.694,1.592){0}{\usebox{\plotpoint}}
\put(1134.02,538.14){\usebox{\plotpoint}}
\put(1154.72,539.67){\usebox{\plotpoint}}
\multiput(1159,540)(20.703,1.479){0}{\usebox{\plotpoint}}
\put(1175.42,541.19){\usebox{\plotpoint}}
\put(1196.12,542.72){\usebox{\plotpoint}}
\multiput(1200,543)(20.694,1.592){0}{\usebox{\plotpoint}}
\put(1216.82,544.27){\usebox{\plotpoint}}
\put(1237.52,545.81){\usebox{\plotpoint}}
\multiput(1240,546)(20.703,1.479){0}{\usebox{\plotpoint}}
\put(1258.22,547.32){\usebox{\plotpoint}}
\put(1278.92,548.85){\usebox{\plotpoint}}
\multiput(1281,549)(20.694,1.592){0}{\usebox{\plotpoint}}
\put(1299.61,550.40){\usebox{\plotpoint}}
\put(1320.31,551.95){\usebox{\plotpoint}}
\multiput(1321,552)(20.703,1.479){0}{\usebox{\plotpoint}}
\put(1341.01,553.46){\usebox{\plotpoint}}
\put(1361.71,554.98){\usebox{\plotpoint}}
\multiput(1362,555)(20.694,1.592){0}{\usebox{\plotpoint}}
\put(1382.41,556.53){\usebox{\plotpoint}}
\multiput(1389,557)(20.694,1.592){0}{\usebox{\plotpoint}}
\put(1403.11,558.08){\usebox{\plotpoint}}
\put(1423.81,559.60){\usebox{\plotpoint}}
\multiput(1429,560)(20.703,1.479){0}{\usebox{\plotpoint}}
\put(1444.51,561.12){\usebox{\plotpoint}}
\put(1456,562){\usebox{\plotpoint}}
\put(120,540){\usebox{\plotpoint}}
\put(120.00,540.00){\usebox{\plotpoint}}
\put(140.58,542.54){\usebox{\plotpoint}}
\multiput(147,543)(20.514,3.156){0}{\usebox{\plotpoint}}
\put(161.17,545.08){\usebox{\plotpoint}}
\put(181.87,546.61){\usebox{\plotpoint}}
\multiput(187,547)(20.547,2.935){0}{\usebox{\plotpoint}}
\put(202.46,549.11){\usebox{\plotpoint}}
\put(223.16,550.65){\usebox{\plotpoint}}
\multiput(228,551)(20.514,3.156){0}{\usebox{\plotpoint}}
\put(243.74,553.20){\usebox{\plotpoint}}
\put(264.44,554.73){\usebox{\plotpoint}}
\multiput(268,555)(20.547,2.935){0}{\usebox{\plotpoint}}
\put(285.03,557.23){\usebox{\plotpoint}}
\put(305.65,559.52){\usebox{\plotpoint}}
\multiput(309,560)(20.694,1.592){0}{\usebox{\plotpoint}}
\put(326.32,561.31){\usebox{\plotpoint}}
\put(346.93,563.68){\usebox{\plotpoint}}
\multiput(349,564)(20.703,1.479){0}{\usebox{\plotpoint}}
\put(367.61,565.35){\usebox{\plotpoint}}
\put(388.21,567.74){\usebox{\plotpoint}}
\multiput(390,568)(20.694,1.592){0}{\usebox{\plotpoint}}
\put(408.85,569.84){\usebox{\plotpoint}}
\put(429.49,571.96){\usebox{\plotpoint}}
\multiput(430,572)(20.703,1.479){0}{\usebox{\plotpoint}}
\put(450.14,573.94){\usebox{\plotpoint}}
\put(470.78,575.98){\usebox{\plotpoint}}
\multiput(471,576)(20.694,1.592){0}{\usebox{\plotpoint}}
\put(491.42,578.06){\usebox{\plotpoint}}
\multiput(498,579)(20.694,1.592){0}{\usebox{\plotpoint}}
\put(512.06,580.15){\usebox{\plotpoint}}
\put(532.66,582.59){\usebox{\plotpoint}}
\multiput(538,583)(20.703,1.479){0}{\usebox{\plotpoint}}
\put(553.35,584.21){\usebox{\plotpoint}}
\put(573.94,586.64){\usebox{\plotpoint}}
\multiput(579,587)(20.694,1.592){0}{\usebox{\plotpoint}}
\put(594.62,588.37){\usebox{\plotpoint}}
\put(615.23,590.71){\usebox{\plotpoint}}
\multiput(619,591)(20.703,1.479){0}{\usebox{\plotpoint}}
\put(635.91,592.45){\usebox{\plotpoint}}
\put(656.52,594.75){\usebox{\plotpoint}}
\multiput(660,595)(20.514,3.156){0}{\usebox{\plotpoint}}
\put(677.10,597.29){\usebox{\plotpoint}}
\put(697.80,598.83){\usebox{\plotpoint}}
\multiput(700,599)(20.547,2.935){0}{\usebox{\plotpoint}}
\put(718.39,601.34){\usebox{\plotpoint}}
\put(739.09,602.86){\usebox{\plotpoint}}
\multiput(741,603)(20.514,3.156){0}{\usebox{\plotpoint}}
\put(759.67,605.41){\usebox{\plotpoint}}
\put(780.26,607.89){\usebox{\plotpoint}}
\multiput(781,608)(20.703,1.479){0}{\usebox{\plotpoint}}
\put(800.96,609.46){\usebox{\plotpoint}}
\put(821.56,611.94){\usebox{\plotpoint}}
\multiput(822,612)(20.694,1.592){0}{\usebox{\plotpoint}}
\put(842.25,613.52){\usebox{\plotpoint}}
\multiput(849,614)(20.514,3.156){0}{\usebox{\plotpoint}}
\put(862.83,616.06){\usebox{\plotpoint}}
\put(883.53,617.58){\usebox{\plotpoint}}
\multiput(889,618)(20.547,2.935){0}{\usebox{\plotpoint}}
\put(904.13,620.09){\usebox{\plotpoint}}
\put(924.76,622.25){\usebox{\plotpoint}}
\multiput(930,623)(20.694,1.592){0}{\usebox{\plotpoint}}
\put(945.42,624.17){\usebox{\plotpoint}}
\put(966.04,626.39){\usebox{\plotpoint}}
\multiput(970,627)(20.703,1.479){0}{\usebox{\plotpoint}}
\put(986.70,628.21){\usebox{\plotpoint}}
\put(1007.32,630.47){\usebox{\plotpoint}}
\multiput(1011,631)(20.694,1.592){0}{\usebox{\plotpoint}}
\put(1027.96,632.57){\usebox{\plotpoint}}
\put(1048.58,634.81){\usebox{\plotpoint}}
\multiput(1051,635)(20.703,1.479){0}{\usebox{\plotpoint}}
\put(1069.25,636.65){\usebox{\plotpoint}}
\put(1089.87,638.85){\usebox{\plotpoint}}
\multiput(1092,639)(20.694,1.592){0}{\usebox{\plotpoint}}
\put(1110.52,640.79){\usebox{\plotpoint}}
\put(1131.16,642.94){\usebox{\plotpoint}}
\multiput(1132,643)(20.703,1.479){0}{\usebox{\plotpoint}}
\put(1151.81,644.89){\usebox{\plotpoint}}
\put(1172.44,646.96){\usebox{\plotpoint}}
\multiput(1173,647)(20.514,3.156){0}{\usebox{\plotpoint}}
\put(1193.03,649.50){\usebox{\plotpoint}}
\multiput(1200,650)(20.694,1.592){0}{\usebox{\plotpoint}}
\put(1213.72,651.10){\usebox{\plotpoint}}
\put(1234.32,653.56){\usebox{\plotpoint}}
\multiput(1240,654)(20.703,1.479){0}{\usebox{\plotpoint}}
\put(1255.01,655.16){\usebox{\plotpoint}}
\put(1275.60,657.61){\usebox{\plotpoint}}
\multiput(1281,658)(20.514,3.156){0}{\usebox{\plotpoint}}
\put(1296.18,660.16){\usebox{\plotpoint}}
\put(1316.88,661.68){\usebox{\plotpoint}}
\multiput(1321,662)(20.547,2.935){0}{\usebox{\plotpoint}}
\put(1337.48,664.19){\usebox{\plotpoint}}
\put(1358.18,665.73){\usebox{\plotpoint}}
\multiput(1362,666)(20.514,3.156){0}{\usebox{\plotpoint}}
\put(1378.76,668.27){\usebox{\plotpoint}}
\put(1399.46,669.80){\usebox{\plotpoint}}
\multiput(1402,670)(20.547,2.935){0}{\usebox{\plotpoint}}
\put(1420.05,672.31){\usebox{\plotpoint}}
\put(1440.66,674.67){\usebox{\plotpoint}}
\multiput(1443,675)(20.694,1.592){0}{\usebox{\plotpoint}}
\put(1456,676){\usebox{\plotpoint}}
\end{picture}

\end	{center}
\vskip 0.15in
\caption{The SU(2) $0^{+}$ mass ($\star$), the 
SU($N\geq 3$)  $0^{++}$ masses ($\bullet$), the
SU($N\geq 3$) $0^{--}$ masses ($\diamond$), and
the average of the $0^{++}$ and $0^{--}$ masses
($\circ$), plotted against $1/N^2$; with the expected
large-$N$ linear dependence shown in each case.}
\label{fig_Scalar}
\end 	{figure}

{\noindent}$\bullet$ In SU(2) there is no $C=-$ sector
and one can ask what the SU(2) $0^+$ mass continues to 
as $N$ increases. This will clearly depend on the dynamics
that produces the ${C=\pm}$ splitting for $N\geq 3$.
For example, if this dynamics simply splits the 
$0^{++}$ and $0^{--}$ 
equally from their naive degenerate masses, then we
would expect the  SU(2) $0^+$ to continue smoothly
to the average of the $0^{++}$ and $0^{--}$ masses.
As another example, if the shift involves just the $0^{--}$,
then the  SU(2) $0^+$ will  continue smoothly to the
$N\geq 3$ $0^{++}$. Conversely, if the shift affects
just the $0^{++}$, then the continuation should be
with the $0^{--}$. In Fig.\ref{fig_Scalar} we plot 
the $0^{++}$ and $0^{--}$ masses, as well as the average 
of the two, as a function of  $1/N^2$. We see that
in all cases the variation with $N$ for $N\geq 3$ can 
be described using just a leading $\propto 1/N^2$
correction. We also note that the  SU(2) $0^+$ mass
extrapolates precisely from the $0^{++}$ masses
while it is inconsistent with a smooth extrapolation 
of the averaged $C=\pm$ states or of the $0^{--}$. 
The same is true for the tensor: the SU(2) $2^+$
mass is a smooth continuation of the SU($N\geq 3$) $2^{\pm+}$
masses, and not of the average of the  $2^{\pm+}$  and
$2^{\pm-}$ masses, or of the $2^{--}$. 
We shall see that this observation provides a tough 
constraint on possible mechanisms for splitting the
$C=+$ and $C=-$ sectors.

{\noindent}$\bullet$ The ${C=\pm}$ splitting appears to 
decrease as the mass increases. To be more specific
we infer from
\be
m_{0^{--}} - m_{0^{++}}
>
m_{0^{--\ast}} - m_{0^{++\ast}}.
\label{eqn_3.1}
\ee 
that states with larger radial quantum number, $n_R$, are 
split less than those with smaller $n_R$. (Recall that
in the flux tube model these lowest excitations are radial
rather than phonon.) Furthermore we infer from
\begin{eqnarray}
m_{0^{--}} - m_{0^{++}} = \ 1.85(26)\surd\sigma 
& > &
m_{2^{--}} - m_{2^{++}} = \ 1.01(38)\surd\sigma 
\nonumber \\
& > &
m_{1^{--}} - m_{1^{++}} = -0.62(65)\surd\sigma 
\label{eqn_3.2}
\end{eqnarray}
(obtained at $N=\infty$) that the magnitude of the splitting also
decreases with increasing phonon number. (Recall that for the 
lightest $J=0,2,1$ glueball states the total phonon number 
in eqn(\ref{eqn_2.3}) is $M=0,2,5$ respectively.) Note that
the decrease we see is even faster when the splitting is
expressed in terms of the average mass. All this 
provides constraints on possible splitting mechanisms.

\subsection{Direct mixing}
\label{subsec_Cdirmix}

Since a flux loop has a direction, $L$ or $R$, it is
convenient to introduce 2-component wave-functions:
\be 
\Psi \equiv
\left( \begin{array}{c}
       \psi_L \\ \psi_R
       \end{array} \right).
\label{eqn_3.5}
\ee 
In this notation, we can write (unnormalized) $C=\pm$ states as
\be 
\Psi_{C=+} = \psi
\left( \begin{array}{c} 1 \\ 0 \end{array} \right)
+ \psi \left( \begin{array}{c} 0 \\ 1 \end{array} \right)
 \ \ ; \ \ 
\Psi_{C=-} = \psi
\left( \begin{array}{c} 1 \\ 0 \end{array} \right)
- \psi \left( \begin{array}{c} 0 \\ 1 \end{array} \right)
\label{eqn_3.7}
\ee 
and eqn(\ref{eqn_2.7}) becomes 
\be
H_\mathit{IP} \Psi = 
\left[ \begin{array}{cc}
       H_L    &  0 \\
       0      &  H_R
       \end{array} \right]
\left( \begin{array}{c}
       \psi_L \\ \psi_R
       \end{array} \right) 
= E \Psi 
\label{eqn_3.8}
\ee
where $H_L\equiv H_R$ is the operator on the LHS of
eqn(\ref{eqn_2.7}).

With the above Hamiltonian there is no mixing between  $L$ and $R$
states and hence no $C=\pm$ splitting. To obtain such a splitting
we need a non-zero probability
for a $L$ state to turn into an $R$ state and vice-versa,
which clearly requires some off-diagonal terms to appear in
the Hamiltonian $H$. So we alter eqn(\ref{eqn_3.8}) 
to define our `direct mixing' Hamiltonian as 
\be
H_\mathit{dir} \Psi = 
\left[ \begin{array}{cc}
       H_L     &  \alpha \\
       \alpha  &  H_R
       \end{array} \right]
\left( \begin{array}{c}
       \psi_L \\ \psi_R
       \end{array} \right) 
= E \Psi 
\label{eqn_3.12}
\ee
where we shall choose to keep $\alpha$ real and constant.
A simple motivation for such a mixing arises from the observation that when the radius of the flux loop is less than the flux tube width, we have something that is no longer a distinct loop, but is rather some kind of ``ball'' which will no longer have any definite orientation.  In a path integral picture, we can think of a path where a loop of orientation $L$ (for example) shrinks into a ball, at which point it loses any memory of its initial orientation, and then expands back out into a loop of either orientation with equal probablility.  This transition will lead to a finite amplitude between $L$ and $R$ loops.

We shall return later to ask how well this model
fits the spectrum. For now we concentrate
on its qualitative predictions, taking the approximation 
$H_L=H_R=M$ where $M$ is the mass of the unmixed state.
Then the energy eigenstates are clearly 
\be  
M_{C=\pm} = M \pm \alpha
\label{eqn_3.17}
\ee 
so that $C=\pm$ states are split equally from their common 
Isgur-Paton value. Three immediate observations follow
from the above:

{\noindent}$\bullet$ For states of approximately equal $\rho$, 
the splitting should be roughly the same 
independent of the
phonon number. This would be the case, for example, for
the lightest $J=0,1,2$ states. However, as we have seen in 
eqn(\ref{eqn_3.2}), the splitting does in fact vary a great deal 
amongst these states.

{\noindent}$\bullet$ It is the average of the $C=+$ and $C=-$ masses
that equals the mass with no mixing. As we decrease $N$
from $N=\infty$ it is this average that should extrapolate
to the SU($N=2$) value of the $0^+$ mass, because the 
Hamiltonian there is the same as $H_L$ or $H_R$. 
(All this up to $O(1/N^2)$ corrections.) However
we have seen in Section~\ref{subsec_Cfeatures} that this is 
not the case: the SU(2) $0^+$ mass equals the SU($N\geq 3$) 
$0^{++}$ mass up to $O(1/N^2)$ corrections. 

{\noindent}$\bullet$ We expect the wavefunction to have a smooth limit
as $N\to\infty$, and so the probability for $\rho$ to be less
than the flux tube radius should also have a non-zero limit. 
Thus the $L\leftrightarrow R$ mixing and the consequent
$C=\pm$ splitting should be non-zero at $N=\infty$, just
as is observed. (This appears to contradict the conventional
statement 
\cite{largeN}
that mixings vanish as $N\to\infty$ but we believe that
the standard arguments do not apply to our kind of `mixing'.)

We might suppose that the first observation becomes irrelevant
when we perform the actual numerical calculations, but unfortunately it
remains an issue. 
Moreover, however we tune the parameters in this model, it 
remains the case that it is the average of the $0^{++}$ and $0^{--}$ 
masses that is predicted to continue smoothly to the  SU(2) 
$0^+$ mass, and, as we have seen, this is contradicted by
the lattice calculations. Thus other, perhaps less straightforward 
mechanisms need to be considered.

\subsection{Adjoint string mixing}
\label{subsec_Cadjmix}

We pointed out in Section~\ref{subsec_strings} that in
SU($N\geq4$) theories there exist extra stable strings and
hence extra states. Since these are just scaled up
versions of our fundamental string spectrum they would
not help in splitting $C=+$ from $C=-$. However there
is another type of string in the theory that
we have not yet considered: the one
that carries adjoint flux. Such a string 
carries no arrow: it is intrinsically $C=+$. 
So any mixing would affect the $C=+$ spectrum:
the $0^{++}$ mass would (probably) be driven down 
while the $0^{--}$ would be left undisturbed.  
But this does not lead to problems with the $0^+$
in SU(2) because the same adjoint string exists in SU(2)
and would drive down the $0^+$ mass there as well.
So we have a mechanism that might at last explain
why the SU(2)  $0^+$ smoothly interpolates onto the
$0^{++}$ for SU($N>2$).

The problem with this mechanism is, of course, that the 
adjoint string can be broken by gluon pair production so it 
is not clear if it makes sense to use a closed adjoint loop 
as the basis for a set of states. However we know that
as $N\to\infty$ the string becomes stable 
\cite{largeN}
so at least for large $N$ it can be used in this
way. Since, as we have seen, lattice studies
\cite{MTd3}
find that the low-lying SU(2) spectrum
differs only by small corrections from SU($N=\infty$),
it is reasonable to assume that the adjoint string
has developed only a modest decay width in SU(2). (As 
indeed seems to be the case
\cite{adjd4}
in 3+1 dimensions.) If so then
the decay time will be long compared to the characteristic
time scale of the low-lying string modes, and we can
safely quantise the string in the Isgur-Paton fashion. 
We will assume that this is so from now on,
although an explicit lattice verification would clearly
be very helpful.

We can use an extension of the formalism in the
previous subsection to derive the Hamiltonian, defining
the wave-function as
\be 
\Psi = \left(  \begin{array}{c}
\psi_+ \\ \psi_- \\ \psi_a \end{array} \right).
\label{eqn_3.50} 
\ee
However since $\psi_a$ always has $C=+$, the Hamiltonian
\be 
H_\mathit{adj} =
\left[ \begin{array}{ccc}
H_+   &   0           &   \alpha_{a} \\
0         & H_-       &   0  \\
\alpha_{a}   & 0  & H_a  \end{array} \right]
\label{eqn_3.52}
\ee
is quite simple: we can clearly reduce it to
a two-component calculation in the $C=+$ sector,
and a simple Isgur-Paton calculation in the $C=-$ sector.
We shall assume that there is no mixing between fundamental
and adjoint loop states that have differing phonon
occupation numbers.

In eqn(\ref{eqn_3.52}) $H_+ = H_-$ is the usual Isgur-Paton
Hamiltonian. $H_a$ will be identical except that the
scale is set by the adjoint string tension, $\sigma_a$,
rather than by the fundamental $\sigma$. It is frequently
speculated that  $\sigma_a$ and $\sigma$ are related
by the ratio of quadratic Casimirs
\be 
{{\sigma_a}\over{\sigma}}
=
{{2N^2}\over{N^2-1}}.
\label{eqn_3.54}
\ee
Lattice calculations in $D=2+1$
\cite{adjd3}
find that for SU(2) $\sigma_a \simeq 2.5\sigma$, which is quite close
to the value of $8/3$ one obtains from eqn(\ref{eqn_3.54}).
Thus we expect states based on the adjoint loop 
to be about a factor of $\sqrt{2.5} \sim 1.5$ heavier than corresponding
states based on the fundamental loop, and so quite massive. Indeed,  
if we assume a simple two body mixing, with $H_+$ replaced
by $M = m_{0^{--}}$ and  $H_a$ by $M_a \simeq  1.5 M$,
and if we choose $\alpha$ so as to obtain the observed $m_{0^{++}}$
mass, we find that the lightest scalar `adjoint' state has
$M_{a,0^{++}} \simeq 1.8 m_{0^{--}}$. This is light enough
to be important by its explicit presence in the spectrum,
in addition to affecting the fundamental states through mixing.

How can the fundamental flux loop mix with an adjoint loop?
Once again our underlying physical picture is that, for
small $\rho$, these loops of string become rather like `balls'
of flux instead, which may reemerge as a loop of different orientation, 
or indeed even of a different type of flux.  We now have the 
following qualitative predictions: 

{\noindent}$\bullet$ As with the direct mixing of 
Section~\ref{subsec_Cdirmix}, we expect the mixing,
and hence the splitting, to be leading order in $N$,
as is observed.

{\noindent}$\bullet$ As remarked above, since the adjoint loop
mixes with just the $C=+$ sector and does so for
all SU($N\geq 2$) gauge theories, we expect the 
SU(2) $0^+$ mass to continue smoothly on to the  
SU($N > 2$)  $0^{++}$ masses, as is observed.

{\noindent}$\bullet$ Whether the splitting decreases with
increasing phonon number $M$ at fixed radial number $n_R$ is
however not clear -- it requires a detailed calculation.

The observed pattern of the $C=\pm$ splitting has proved
very constraining on possible dynamics, but the model
in which the fundamental loop mixes with an adjoint loop
appears to have the right qualitative features.
In the next section we shall see how well it can do
at the quantitative level.

\section{Fitting the lattice spectrum}
\label{sec_fits}

We have performed a large variety of comparisons with the 
lattice spectrum. We have used the different $C=\pm$ 
splitting mechanisms described in this paper; 
we have fitted to all the lattice masses or just
to a more reliable subset as discussed above; we have used
the actual errors in the $\chi^2$ or expanded errors
as described below; we have used explicit short distance 
fudge factors of various kinds or a $\rho$-dependent 
string tension as described earlier, or indeed no suppression
at all. The reader will be relieved to learn that we do not
intend to describe this very large 
number of model fits here but will rather focus on
two of the most relevant examples. Some different model fits 
can be found in
\cite{lat97}
and others will appear in
\cite{RJthesis}.

The generalised flux tube model typically has two or three
unknown parameters. First there is $\gamma$, the 
(renormalised) curvature/elasticity.
Secondly there is the mixing parameter characterised
by a strength $\alpha_0$. There is the parameter $f$ which
characterises the short-distance cut-off imposed in
the Isgur-Paton model. Alternatively we can use
$\sigma_{eff}(\rho)$ instead of $\sigma$ at short distances 
and this enables us to do without the parameter $f$.
We can then solve the model for various values of these
parameters and find which fits the lattice spectrum the best.

In this section we will describe what we find when
we fit the lattice data with adjoint string mixing,
as described in Section~\ref{subsec_Cadjmix}, or with
direct mixing, as described in Section~\ref{subsec_Cdirmix}.
In both cases we shall use a $\rho$-dependent string tension,
as described in Section~\ref{subsec_f}, to embody the
short distance corrections to the flux tube picture, in
a way that introduces no new parameters. In our physical 
picture of the mixing, we see it as occuring at small $\rho$
and so we choose the mixing term in 
eqns(\ref{eqn_3.12}, \ref{eqn_3.52}) to satisfy
\be 
\alpha(\rho) = 
\alpha \,
\biggl(1 - {{\sigma_{eff}(\rho)}\over {\sigma_{eff}(\infty)}}
\biggr)
\label{eqn_4.20} 
\ee
(where we now drop the subscript on $\alpha_a$). In the case
of adjoint mixing we determine the adjoint string tension
from eqn(\ref{eqn_3.54}). Thus there are two
parameters to be fitted: $\gamma = \gamma_E - 13/12$, where
$\gamma_E$ is the string curvature 
described in Section~\ref{subsec_gamma}, and $\alpha$,
the mixing strength.

There are however some problems with just taking the mass 
spectrum as given in Table~\ref{table_mlattice}
and doing a least-$\chi^2$ fit on the model parameters.
First we have already seen that the model prediction is 
going to be far from the lattice $0^{-+}$, so including
it in the fit might badly distort the final `best fit'.
Moreover there are theoretical reasons for thinking that
the lattice $0^{-+}$ spin assignment might be mistaken and that 
this is actually a $4^{-+}$. (And, similarly, that the lattice 
$0^{+-}$ is really a $4^{+-}$.) To a lesser extent similar
questions arise with the purported $J=1$ states.
We shall therefore exclude these states
from the fit but will instead quote the values for
these masses, as predicted by the best fit
to the other states.

A second problem is that the most accurate masses are
for the lightest states. Having the smallest errors
these will provide the most important component of the
$\chi^2$ function that determines the best fit.
On the other hand the lightest states possess the
smallest radius $\rho$ and so we expect the flux loop 
model to suffer the largest corrections for such states.
Thus we might not want them to dominate, and so possibly
distort, the best fit. 

This second problem has no unambiguous resolution. The fact
is that we know that the model is a simple approximation 
that can at best incorporate only the essential features of
glueball structure. The most that we can expect is that
it roughly reproduces the spectrum and whether it can do
so is what we want to learn when we fit the model to the
data. To do this we need to embody what we mean by `roughly'
in some specific way into the fitting procedure. In order not to be unduly 
biased by the very small errors on the lightest masses, we 
have enhanced all the errors by 5\% of the mass, and add it in quadrature with 
the statistical error. In practice the best fits turn
out to be  similar whether we perform such an
enhancement or not.  
That is to say, we are saying that we want to know if
the model can fit the masses within  $\pm$5\%. Of course
the best fit might do better than that; but at least it
will not be driven by the very small errors on the
lightest masses for which it is probably least reliable.

In each case we obtain the predicted spectrum by solving
the coupled set of differential equations, using the
numerical technique detailed in
\cite{RJthesis},
on a grid in parameter space, and then we perform a final simplex 
minimization starting at the most favorable grid point to find
the best $\chi^2$ fit. (With just two parameters such a crude 
approach works reasonably well.)  The best fit is determined
using the lightest three states in the $0^{++}$ and $0^{--}$ sectors,
and the two lightest states in the $2^{\pm +}$ and $2^{\pm -}$ sectors.

\begin{figure}[p]
\begin{center}
\epsfig{figure=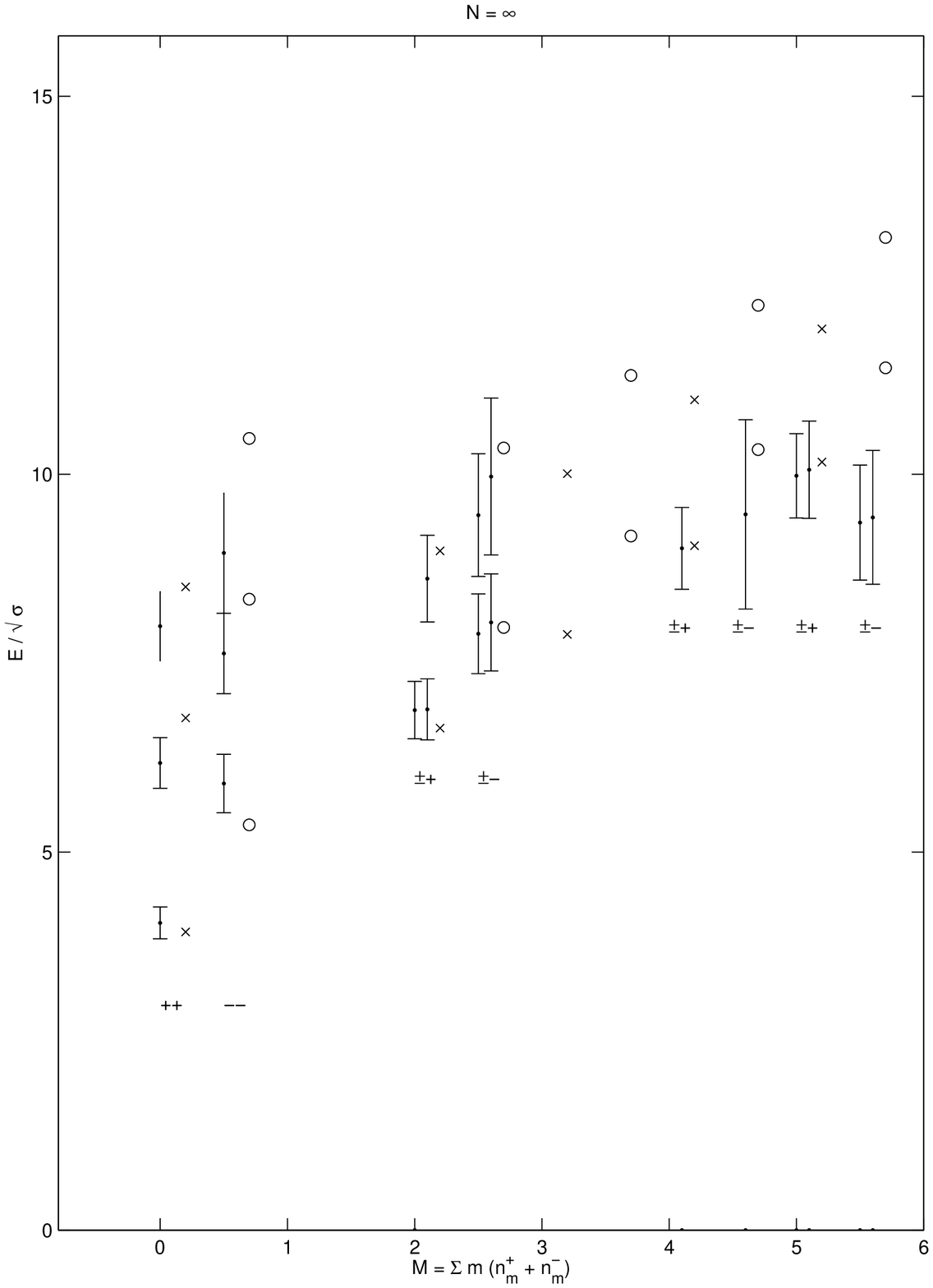, height=.8\textheight} 
\end{center}
\vskip 0.15in
\caption[]{\label{fig_adjfitN}
{ The spectrum for the adjoint mixing mechanism compared to
lattice data at $N=\infty$.  The x-axis gives the total phonon 
number M.  The $0^{-+}$ and $0^{+-}$ are compared with M=4, and states 
with J=1 are compared with M=5.}}
\end{figure}

\begin{figure}[p]
\begin{center}
\epsfig{figure=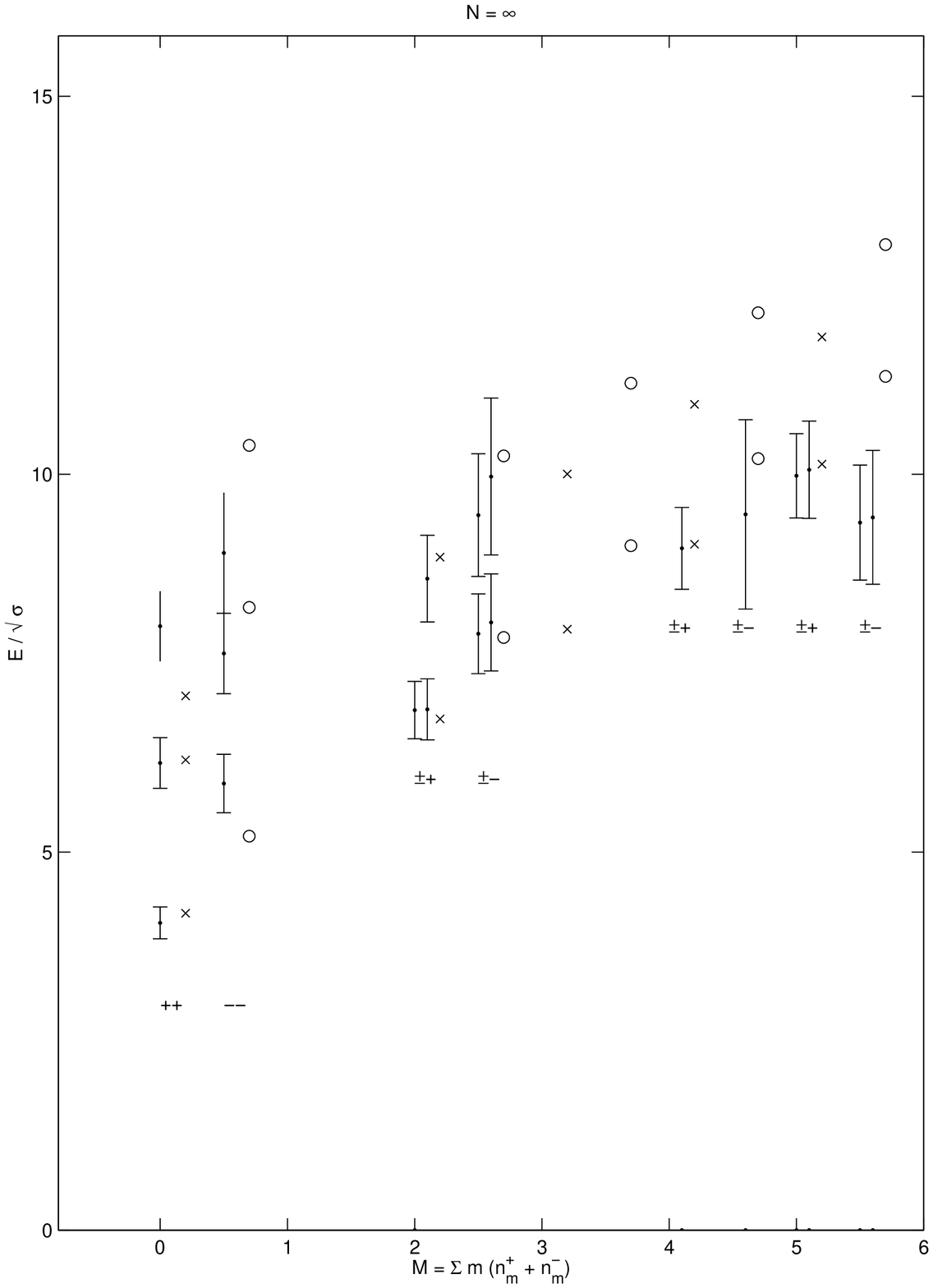, height=.8\textheight} 
\end{center}
\vskip 0.15in
\caption[]{\label{fig_dirfitN}
{ The spectrum for the direct mixing mechanism compared to
lattice data at $N=\infty$.  The x-axis gives the total phonon 
number M.  The $0^{-+}$ and $0^{+-}$ are compared with M=4, and states 
with J=1 are compared with M=5.}}
\end{figure}

We have remarked that
there are reasons to think that these simple models
should work best in the $N\to\infty$ limit, and so
we display the results of fitting to the lattice spectrum
as extrapolated to $N\to\infty$.  A more thorough display of the
results for all N may be found in
\cite{RJthesis}.
In Table~\ref{table_adjfitN} we show the mass spectra 
for the adjoint mixing mechanism, and in Table~\ref{table_adjpar} we present the values of the fitted parameters, for all values of N.  In 
Figure~\ref{fig_adjfitN} we display the spectrum at $N=\infty$.
The corresponding spectra
for the direct mixing mechanism are shown in
Table~\ref{table_dirfitN}, and in Table~\ref{table_dirpar}
are the parameters for the direct mechanism.  Figure~\ref{fig_dirfitN}
displays the direct mixing spectrum at $N=\infty$.

We see from the Figures that the best fits, whether obtained
using the direct or the adjoint mixing mechanisms,
are of an overall reasonable quality. 
A closer examination of the detailed spectra does 
however show that both models have some minor difficulties. 
The direct mechanism has problems with the excitations in the
J=0 sector:  by the time the third excitation is reached, which
has a greater average $\rho$ than the ground state, the conjugate
mixing drives the two model states much farther apart than they ought to be. 
The way that the adjoint mixing model avoids this
problem is that the mass of the lightest adjoint loop state 
is naturally close to that of the  $0^{++\star}$, 
because $\sqrt{\sigma_a/\sigma_f} \sim 1.5$.
Thus, after mixing, the  $0^{++\star}$ can be largely
an adjoint loop, and the  $0^{++\star\star}$ can
then be (largely) the first fundamental loop excitation.
However, the direct mixing mechanism does have numerically better 
values of $\chi^2$, and thus we ought to consider it more likely than
the adjoint mixing mechanism. 
This suggests that a flux tube model that combines
both direct and adjoint mixing should be able to do
much better than either model alone. Combining these
models is quite natural; our picture for the way
an oriented fundamental loop may evolve into an
adjoint loop (through contracting into a small 
`disoriented' ball) is precisely the way we saw
the direct mixing between fundamental loops of
opposite orientation proceeding. This picture will,
in general, require two mixing parameters. While
it is interesting to explore these ideas
\cite{RJthesis}
the analysis would clearly benefit from improved
lattice calculations where a larger number of excited
states are accurately determined.

Returning to the best fit parameters listed in 
Tables~\ref{table_adjpar} and ~\ref{table_dirpar},
we observe that all our fits require $\gamma$ to
be positive or very close to zero. Ignoring the values for SU(2), 
there is a trend in the values of $\gamma$ consistent with a
$1\over N^2$ relationship; see 
\cite{RJthesis}
for more details.
Taken together with eqn(\ref{eqn_2.8}) 
this tells us that the observed mass spectra do indeed
require  a non-zero curvature (elasticity) term,
$\gamma_E \in [0.5,1.0]$, in the effective string model
for the confining flux tube. 

From either Figure~\ref{fig_adjfitN} or Figure~\ref{fig_dirfitN} we see
that the conventional spin assignments of the heavier states are called into
question.  The model consistently puts the $4^{-+}$ and the $4^{+-}$ states
at masses corresponding to lattice states with spin 0.  These states do not
contribute to the $\chi^2$, and so their masses are purely predicitve--the 
agreement is remarkable.  Furthermore, while the lattice states $1^{\pm+}$
agree with the model's predictions for the lightest state with $M=5$ (which
is the smallest value of M giving spin 1), the states conventionally labelled
$1^{\pm-}$ are at the mass predicted for states with $M=3$.  These states 
are also related by a spin ambiguity of {\it{modulo}}4 on a cubic lattice, 
and so we must also investigate these assignments in the future as well.

Finally we remark that
the features we have described are not only robust
against the detailed fitting procedures used but
much the same conclusions are obtained if we replace
the $\rho$ dependence of the string tension with
a short distance modification of the kind 
shown in eqn(\ref{eqn_2.6}).

\section{Conclusions}
\label{sec_conc}

In this paper we set out to test the idea that glueballs
are quantised closed strings of colour-electric flux. 
Such a picture arises naturally in linearly confining theories, 
such as SU($N$) gauge theories in 2+1 and 3+1 dimensions, 
where distant fundamental charges are connected by flux tubes.

We started out with the specific dynamical framework
of the Isgur-Paton flux tube model 
\cite{IsgPat},
in which the excitations of the closed flux loop are
either radial or phonon-like, 
and we confronted its mass spectrum with the rather accurate 
mass spectra available in D=2+1 SU($N$) gauge theories
\cite{MTd3}.
As $N\to\infty$ the gauge theory simplifies in ways which
bring it closer to some of the model's assumptions, e.g.
the neglect of decays, and so one can argue that a comparison
in this limit makes particular sense. If we express the 
observed masses in units of the observed string tension, 
then the model's predictions for these dimensionless ratios 
involve no free parameters at all. 
We found that these predictions were, for the most part,
quite remarkably good in the $C=+$ sector of states;
and for the $C=-$ sector they embodied the main qualitative 
features even if the quantitative comparison was less good.

Of course, when 
the flux tube radius, $\rho$, is smaller than the flux tube width
the picture must break down,  embodied 
in the model by suppressing the potential energy 
below $\rho \sim 1/f$, where $f$ is a parameter that needs to
be determined but which we expect to be $O(\surd\sigma)$. In
this paper we described other ways of including such a cut-off;
in particular through a dependence of the string tension
on $\rho$. By taking $\sigma(\rho)$ from calculations
in the literature
\cite{KST99},
one can avoid having the additional parameter $f$
to fit. We then pointed out that one should in general
include a string curvature term, which for a closed loop will 
make a  contribution $\gamma_E/\rho$ to the effective potential
that is of the same form as the Casimir string energy. 
This introduces a parameter $\gamma$ that needs to be determined. 
Finally we noted that the  $C=\pm$ degeneracy in the model mass
spectrum is contradicted by the splittings seen in the lattice
spectrum and we were compelled to consider dynamical mechanisms
that might reproduce these splittings. Such mechanisms 
typically involve a mixing parameter $\alpha$
that also needs to be determined.
In searching for the best fit to the observed spectrum,
we found that whatever mixing mechanism we used we 
invariably required a substantial positive string curvature 
contribution, $\gamma_E \in [0.5,1]$. 

The qualitative features of the observed $C=\pm$ splittings
proved to be very constraining. The only mechanism that we 
were able to construct that was natural, simple and had
the right qualitative behaviour, involved adding to 
the basic flux tube model a sector of states built on closed 
loops of adjoint flux. These are intrinsically $C=+$ and we
introduced a mixing between these states and the $C=+$
states built on the fundamental loop. We pictured the mixing 
as arising at small $\rho$ where a closed flux tube becomes 
a flux-less `ball', and we conjectured that this kind
of mixing may be leading order in $1/N$, as required by 
the lattice spectrum. 
The implication of these calculations and also
calculations with other mixing mechanisms, such as a direct 
mixing between fundamental loops of opposite orientation, 
was that the simultaneous presence of adjoint loop and
direct mixing was likely to be much more successful in
quantitatively reproducing the observed spectrum.

However even these best
fits always left us with one very large discrepancy:
the $0^{-+}$ state. In the model this is a highly excited state
(involving eight phonon units) and is predicted to
be much heavier than the lattice $0^{-+}$. This is
a robust result of the model: the splitting of
the $0^{++}$ and $2^{++}$ states, which is two
phonon units, is what essentially determines the 
$0^{-+}$ mass. On the other hand the predicted $4^{-+}$ 
mass is very close to the lattice  $0^{-+}$ mass
(and also for the model's $4^{+-}$ and lattice $0^{+-}$).
This suggests that the lattice calculation may have
mislabelled this state; after all one cannot distinguish 
$J=0$ from $J=4$ by the rotational symmetries of a
square lattice. This has provoked lattice calculations
that are attempting to resolve this rather basic question
\cite{lat98,rot01}.
(For a detailed exposition of the problem, as well as a 
means for its solution, see
\cite{RJthesis}.)

If there are states built on the adjoint loop they will be
about half as heavy again as the corresponding states built on the 
fundamental loop and so only the very lightest are likely
to be present in the currently available lattice spectrum.
Moreover, since the adjoint loop can break, these states 
may have large decay widths, their masses may be
shifted from their naive values, and perhaps only
the lightest states will actually exist. However, with
a modest improvement in the quality of the lattice
calculations, one could search for their presence.
The same improvement would allow us to search for
degeneracies between states of differing $J$ but with
the same number of phonon `units'. For example
there will be an excited $0^{++}$ with phonon content
$n_{m=2}^+ = n_{m=2}^- = 1$ that should be degenerate
with a $4^{++}$ with phonon content $n_{m=2}^+ = 2, \,
n_{m=2}^- = 0$. Such (near) degeneracies provide
a characterisitic pattern that would test the
general dynamics of the Isgur-Paton flux tube model.

In SU($N\geq 4$) gauge theories there are additional strings
than the fundamental which will be stable if their string
tensions are low enough. In D=3+1 this is expected to be the 
case on theoretical grounds
\cite{strassler},
and indeed is known to be the case for SU(4)
\cite{blmtsu4,winsu4}.
Thus the string model predicts that, if we neglect mixings and 
decays, the observed mass spectrum will contain towers of states 
that are exact scaled-up replicas of the spectrum arising from
the fundamental string, and that the number of these `towers'
of states will grow as $N\to\infty$. This is a dramatic
and robust prediction of the basic flux tube picture
which can and needs to be investigated by lattice
calculations.

We have seen that the kind of exercise undertaken in
this paper, testing a model against lattice
calculations, has a fruitful impact in both directions.
We have been forced to generalise the model in ways
that, in retrospect, are entirely natural. And the
model points to both potential weaknesses in the lattice
calculations and motivates specific further calculations
that promise to be very informative however they turn out.

%
%
\section*{Acknowledgments}
We are grateful to Jack Paton for valuable discussions
during the course of this work. One of us (RWJ) would
like to thank the Rhodes Trust for financial support.

\vfill \eject


\section*{Appendix}

In this Appendix we briefly describe how we derive the canonical
variables for the Isgur-Paton model. First we discuss the traditional
model's equation, and then we describe how we generalise
the formalism to accommodate an effective string tension that
varies with the loop radius.

\vspace{0.45cm}
\noindent{\bf{1. The Isgur-Paton Hamiltonian}}
\vspace{0.35cm}

We have a circular loop of mass $\mu = 2\pi \rho$ which moves in
an effective potential provided by the phonon modes etc. as
discussed earlier in this paper. Thus the kinetic and potential
terms are: 
\be 
T = \frac{p_\rho^2}{4\pi\sigma \rho} \ \ \ ; \ \ \ 
V = 2\pi\sigma \rho + F(\rho) \frac{M+\gamma}{\rho}.
\label{eqn_A1}
\ee
where  $F(\rho)$ is the usual short distance fudge-factor, which we
shall now  set to unity for convenience. Under the transcription
$p_\rho \rightarrow i\hbar\frac{\pa}{\pa \rho}$ we have 
(with $\hbar = 1$)
\be
T = \frac{p_\rho^2}{2\mu} 
\rightarrow 
\frac{-1}{4\pi\sigma}
\frac{\frac{\pa}{\pa \rho}\frac{\pa}{\pa \rho}}{\rho} . 
\label{eqn_A3}
\ee
We want canonical variables $\xi,P_\xi$ such that $T$ is 
quadratic in $P_\xi$. In passing from the above classical system 
to the quantum system there is an ambiguity how to order the
various terms in $T$. We choose
$P_\rho = \frac{\frac{\pa}{\pa \rho}}{\rho^\half}$
so that $T \propto P^2_\rho$. That is to say we choose
\be 
\xi = \rho^{3\over 2},
\label{eqn_A5}
\ee
so that 
\be 
T  \rightarrow
\frac{-1}{4\pi\sigma}\left( {3\over 2} \right)^2 
\frac{\pa^2}{\pa \xi^2} \\
 =
\frac{-9}{16\pi\sigma}\frac{\pa^2}{\pa \xi^2}
\label{eqn_A7}
\ee
and we arrive at eqn(\ref{eqn_2.7}).

\vspace{0.45cm}
\noindent{\bf{2. Generalising to ${\bf \sigma(\rho)}$}}
\vspace{0.35cm}

If the vacuum of non-Abelian gauge theories is a (type-II)
dual superconducter, then  flux tubes will arise through a 
dual Meissner effect. Thus it is interesting and relevant 
to ask how closed flux tubes behave as the radius is varied
in the  Ginzburg-Landau theory of type-II superconductors.
As pointed out in
\cite{KST99}
what happens is that the effective string tension
$\sigma_\mathrm{eff}(\rho)$ vanishes as $\rho\to 0$, roughly as 
in eqn(\ref{eqn_2.46}). If we replace $\sigma$ in the 
Isgur-Paton model by such a $\sigma_\mathrm{eff}(\rho)$ we
will be providing the model with a short-distance cut-off
that is both natural and employs no free parameters.
The price for this is some complication in the quantisation
of the model. However, as we shall now see, this is a problem
that can be overcome.

As we saw earlier, the tricky term in quantizing the Isgur-Paton
model is the kinetic energy, so that is where we will begin.
Abusing notation slightly,
\be
T = \frac{p_\rho^2}{2 \mu} \\
 \rightarrow \frac{-1}{4 \pi} 
\left( \frac{\frac{\pa}{\pa \rho}}
{\sqrt{\sigma_\mathrm{eff}(\rho) \rho}}\right)^2 , 
\label{eqn_A30}
\ee
where $\sigma_\mathrm{eff}(\rho)$ is now included in the 
$\rho$ dependence of the kinetic energy.  
If $\sigma_\mathrm{eff}$ were a constant, we would get
$T$ as in eqn(\ref{eqn_A7}). Now we must find a canonical variable
$\xi$ such that
\be
\frac{\frac{d\xi}{d\rho}\frac{d}{d\xi}}
{\sqrt{\sigma_\mathrm{eff}(\rho) \rho}} 
= \frac{d}{d\xi} 
\label{eqn_A32}
\ee
or
\be
d\xi = \sigma_\mathrm{eff}^\half(\rho) \rho^\half d\rho
\label{eqn_A34}
\ee
Substituting the function in 
eqn(\ref{eqn_2.46}) for $\sigma_\mathrm{eff}$, we find
\be
\xi = \sigma^\half 
\int d\rho \rho^\half \left( 1 - e^{-1.72 \rho} \right)^\half . 
\label{eqn_A36}
\ee
where the integration constant has been determined by
the boundary condition $\xi(0)=0$.
This integral cannot be put into closed form, yet we need the 
inverse function explicitly so as to substitute it into the 
potential $V(\rho(\xi))$.  While it would be nice to include 
explicitly the function from 
\cite{KST99},
any function which approximates it reasonably well will suffice.
So rather than using eqn(\ref{eqn_2.46}) we shall use a less 
natural form, but one that will suit our manipulations better.
Now, we need a function for $\sigma_\mathrm{eff}(\rho)$
which goes to zero as $\rho \rightarrow 0$ and which approaches 
$\sigma$ as $\rho \rightarrow \infty$, 
and which is integrable with respect to the 
measure $\rho^\half d\rho$. As an example consider
\be
\sigma_\mathrm{eff}(\rho) = \sigma
\left( 1 - e^{-f \rho^{3\over 2}} \right)^2
\label{eqn_A38}
\ee
which, with a suitable value for $f$, will crudely approximate 
the dependence found in
\cite{KST99}.
Performing the integral, and imposing $\xi(0)=0$, we have
\be
{\xi} = \sigma^\half
\int d\rho \rho^\half \left( 1 - e^{-f \rho^{3\over 2}} \right) 
= {2\over 3} \sigma^\half
\left( \rho^{3\over 2} - {1 \over f} + 
{1 \over f} e^{-f \rho^{3\over 2}}
\right).
\label{eqn_A40}
\ee
We cannot find the functional inverse explicitly, but as we are 
ultimately working with a discrete set $\{\xi_j\}$ related to $\{t_j\}$, 
we can simply solve the equation numerically to give $\rho_j$
which we can put into the potential $V = V(\rho(\xi_j))$.
(In practice we have used 
$\sigma_\mathrm{eff} =\sigma ( 1 - e^{-f \rho})^2$
which leads to a slightly more complex relation than the above.)
Returning to the kinetic energy,
\be
T \rightarrow 
\frac{-1}{4 \pi} \frac{\pa^2}{\pa \xi^2} ,
\label{eqn_A44}
\ee
and we proceed with the numerical solution of the discretised
equation as described earlier. 

\vfill

\eject

\begin{table}
\begin{center}
\begin{tabular}{|l||l|l|l|l||l|}\hline
\multicolumn{6}{|c|}{$m_G/\surd\sigma$} \\ \hline
state & SU(2) & SU(3) & SU(4) & SU(5) & SU($\infty$) \\ \hline
$0^{++}$         & 4.718(43) & 4.329(41) & 4.236(50) & 4.184(55) & 4.065(55) \\
$0^{++\ast}$     & 6.83(10)  & 6.52(9)  & 6.38(13) & 6.20(13) & 6.18(13) \\
$0^{++\ast\ast}$ & 8.15(15)  & 8.23(17) & 8.05(22) & 7.85(22) & 7.99(22) \\ 
$0^{--}$         &           & 6.48(9)  & 6.271(95) & 6.03(18) & 5.91(25)  \\ 
$0^{--\ast}$     &           & 8.15(16) & 7.86(20) & 7.87(25)  & 7.63(37) \\ 
$0^{--\ast\ast}$ &           & 9.81(26) & 9.21(30) & 9.51(41)  & 8.96(65) \\
$0^{-+}$         & 9.95(32)  & 9.30(25) & 9.31(28) & 9.19(29)  & 9.02(30) \\
$0^{+-}$         &           & 10.52(28) & 10.35(50) & 9.43(75) & 9.47(116)  \\
$2^{++}$         & 7.82(14)  & 7.13(12) & 7.15(13) & 7.19(20) & 6.88(16) \\
$2^{++\ast}$     &           & --         & 8.51(20) & 8.59(18) & -- \\
$2^{-+}$         & 7.86(14)  & 7.36(11) & 6.86(18) & 7.18(16) & 6.89(21) \\
$2^{-+\ast}$     &           & 8.80(20) & 8.75(28) & 8.67(24) & 8.62(38) \\
$2^{--}$         &           & 8.75(17) & 8.22(32) & 8.24(21) & 7.89(35) \\
$2^{--\ast}$     &           & 10.31(27) & 9.91(41) & 9.79(45) & 9.46(66)  \\
$2^{+-}$         &           & 8.38(21)  & 8.33(25) & 8.02(40) & 8.04(50)  \\
$2^{+-\ast}$     &           & 10.51(30) & 10.64(60) & 9.97(55) & 9.97(91) \\
$1^{++}$         & 10.42(34) & 10.22(24) & 9.91(36) & 10.26(50) & 9.98(25) \\
$1^{-+}$         & 11.13(42) & 10.19(27) & 10.85(55)& 10.28(34) & 10.06(40) \\
$1^{--}$         &           & 9.86(23)  & 9.50(35) & 9.65(40) & 9.36(60) \\
$1^{+-}$         &           & 10.41(36) & 9.70(45) & 9.93(44) & 9.43(75) \\ 
\hline
\end{tabular}
\caption{\label{table_mlattice}
Glueball masses in units of the string tension; in the continuum
limit
\cite{MTd3}.
The SU($\infty$) values are obtained by extrapolating 
the SU($N\leq 5$) values with an $O(1/N^2)$ correction. }
\end{center}
\end{table}

\begin{table}
\begin{center}
\begin{tabular}{|l|l|l|}\hline
\multicolumn{3}{|c|}{$m_G/\surd\sigma$} \\ \hline
 $J^{PC}$ &  SU($\infty$) & IP model \\ \hline
$0^{++}$         & 4.065(55) & 3.12  \\
$0^{++\ast}$     & 6.18(13)  & 6.46  \\
$0^{++\ast\ast}$ & 7.99(22)  & 8.72  \\
$2^{\pm +}$      & 6.88(16)  & 6.79  \\ 
$2^{\pm +\ast}$  & 8.62(38)  & 9.06  \\
$0^{-+}$         & 9.02(30)  & 13.86 \\
$4^{\pm +}$      &  --       & 9.64  \\
$1^{\pm +}$      & 10.00(25) & 10.84 \\
$3^{\pm +}$      &  --       & 8.30  \\ \hline
$0^{--}$         & 5.91(25)  & 3.12  \\
$0^{--\ast}$     & 7.63(37)  & 6.46  \\
$0^{--\ast\ast}$ & 8.96(65)  & 8.72  \\
$2^{\pm -}$      & 7.94(35)  & 6.79  \\ 
$2^{\pm -\ast}$  & 9.62(66)  & 9.06  \\
$0^{+-}$         & 9.47(116) & 13.86 \\
$4^{\pm -}$      &  --       & 9.64  \\
$1^{\pm -}$      & 9.38(60)  & 10.84 \\
$3^{\pm -}$      &  --       & 8.30  \\
\hline
\end{tabular}
\caption{\label{table_tepmor}
Glueball masses in units of the string tension. Predictions
of the simple no-parameter Isgur-Paton flux tube model 
compared to the actual spectrum of the SU($N=\infty$) theory.}
\end{center}
\end{table}

\newpage

\begin{table}
\begin{center}
\begin{tabular}{|l|l|l|l|}\hline
\multicolumn{4}{|c|}{adjoint loop mixing} \\ \hline
 group & $\alpha$ &   $\gamma$ & $\chi^2_5$/d.o.f
 \\  \hline
 SU(2) &  0.86$\pm$0.65 & -0.19$\pm$0.07  & 0.47  \\
 SU(3) &  4.5$\pm$0.60 & 0.57$\pm$0.09  & 1.1 \\
 SU(4) &  4.2$\pm$0.66 & 0.42$\pm$0.08  & 1.2  \\
 SU(5) &  3.8$\pm$0.84 & 0.25$\pm$0.08  & 1.4 \\
 SU($\infty$) & 3.3$\pm$0.8 & 0.07$\pm$0.07 & 0.99  \\
\hline
\end{tabular}
\caption{\label{table_adjpar}
Best fit parameters of the adjoint mixing model.}
\end{center}
\end{table}                                                

\begin{table}
\begin{center}
\begin{tabular}{|l||l|l|l|l|l|}\hline
\multicolumn{6}{|c|}{$m_G/\surd\sigma$} \\ \hline
state & SU(2) & SU(3) & SU(4) & SU(5) & SU($\infty$) \\ \hline
$0^{++}$         & 4.87 & 4.29 & 4.09 & 4.04	& 3.94\\
$0^{++\ast}$     & 6.84 & 7.03 & 6.86 & 6.86	& 6.78 \\ 
$0^{++\ast\ast}$ & 8.05 & 9.05 & 8.88 & 8.81	& 8.51 \\ 
$0^{--}$         & 	& 6.04 & 5.83 & 5.60	& 5.36 \\ 
$0^{--\ast}$     & 	& 8.86 & 8.70 & 8.53	& 8.35 \\ 
$0^{--\ast\ast}$ & 	& 10.90 & 10.77 & 10.63	& 10.48 \\ 
$0^{-+}$         & 14.06 & 14.13 & 14.00 & 13.93&  13.78 \\ 
$0^{+-}$         & 	& 14.21 & 14.07 & 13.98 &  13.82 \\ 
$4^{-+}$         & 11.05  & 10.25  & 10.13  & 10.17& 10.16 \\ 
$4^{+-}$         & 	& 11.93 & 11.77 & 11.59    & 11.41 \\ 
$2^{\pm+}$       & 7.55 & 6.87 & 6.70 & 6.70 	& 6.64 \\ 
$2^{\pm+\ast}$   & 9.96 & 9.11 & 8.97 & 9.01	& 8.99 \\ 
$2^{\pm-}$       & 	& 8.59 & 8.40 & 8.19	& 7.97 \\ 
$2^{\pm-\ast}$   & 	& 10.83 & 10.68 & 10.52 & 10.35 \\ 
$1^{\pm+}$       & 9.95 & 9.19  & 9.05  & 9.07  & 9.05 \\ 
$1^{\pm-}$       & 	& 10.87 & 10.70 & 10.52 & 10.33 \\ 
$3^{\pm+}$       & 8.78 & 8.06 & 7.91 & 7.92 	& 7.88 \\ 
$3^{\pm-}$       & 	& 9.76 & 9.58 & 9.39 	& 9.18 \\ 
\hline
\end{tabular}
\caption{\label{table_adjfitN}
Best fit spectrum of the adjoint-mixing flux tube model to the 
SU($N$) glueball masses, in units of the string tension.}
\end{center}
\end{table}

\begin{table}
\begin{center}
\begin{tabular}{|l|l|l|l|}\hline
\multicolumn{4}{|c|}{direct mixing} \\ \hline
 group & $\alpha$ &   $\gamma$ & $\chi^2_5$/d.o.f \\  \hline
 SU(2) &  --          & -0.42$\pm$0.12   & 1.6  \\
 SU(3) & 2.17$\pm$0.6 & 0.57$\pm$0.10   & 1.0  \\
 SU(4) & 2.15$\pm$0.6 & 0.41$\pm$0.10   & 1.1  \\
 SU(5) & 1.98$\pm$0.6 & 0.28$\pm$0.10   & 1.3  \\
 SU($\infty$) & 1.36$\pm$0.6 & -.03$\pm$0.11 & .65  \\
\hline
\end{tabular}
\caption{\label{table_dirpar}
Best fit parameters of the direct mixing model.}
\end{center}
\end{table}

\begin{table}
\begin{center}
\begin{tabular}{|l||l|l|l|l|l|}\hline
\multicolumn{6}{|c|}{$m_G/\surd\sigma$} \\ \hline
state 		& SU(2) & SU(3) & SU(4) & SU(5) & SU($\infty$) \\ \hline
$0^{++}$         & 4.67 & 4.37  & 4.19  & 4.13  & 4.19 \\
$0^{++\ast}$     & 7.84 & 6.98  & 6.84  & 6.84  & 6.22 \\ 
$0^{++\ast\ast}$ & 9.77 & 7.66  & 7.43  & 7.11  & 7.07 \\ 
$0^{--}$         &      & 6.03  & 5.82  & 5.64  & 5.21 \\ 
$0^{--\ast}$     &      & 8.86  & 8.69  & 8.56  & 8.24 \\ 
$0^{--\ast\ast}$ &      & 10.90 & 10.76 & 10.65 & 10.38 \\ 
$0^{-+}$         & 13.86 & 13.66 & 13.52 & 13.48 & 13.32 \\ 
$0^{+-}$         &       & 14.43 & 14.25 & 14.15 & 14.01 \\ 
$4^{-+}$         & 10.13 & 10.04 & 9.90  & 9.91  & 10.13 \\ 
$4^{+-}$         &      & 11.92 & 11.76 & 11.62 & 11.29 \\ 
$2^{\pm+}$       & 7.35 & 6.83  & 6.67  & 6.64  & 6.76 \\ 
$2^{\pm+\ast}$   & 9.86 & 8.92  & 8.79  & 8.80  & 8.90 \\ 
$2^{\pm-}$       &      & 8.58  & 8.39  & 8.22  & 7.84 \\ 
$2^{\pm-\ast}$   &      & 10.83 & 10.68 & 10.54 & 10.24 \\ 
$1^{\pm+}$       & 10.88 & 9.03 & 8.89  & 8.88  & 9.07 \\ 
$1^{\pm-}$       &      & 10.87 & 10.70 & 10.55 & 10.21 \\ 
$3^{\pm+}$       & 8.59 & 7.96  & 7.80  & 7.79  & 7.95 \\ 
$3^{\pm-}$       &      & 9.76  & 9.58  & 9.42  & 9.06 \\ 
\hline
\end{tabular}
\caption{\label{table_dirfitN}
Best fit spectrum of the direct-mixing flux tube model to the 
SU($N$) glueball masses, in units of the string tension.}
\end{center}
\end{table}

\newpage

\end{document}